\input jytex.tex   
\typesize=10pt \magnification=1200 \baselineskip17truept
\footnotenumstyle{arabic} \hsize=6truein\vsize=8.5truein
\sectionnumstyle{blank}
\chapternumstyle{blank}
\chapternum=1
\sectionnum=1
\pagenum=0

\def\begintitle{\pagenumstyle{blank}\parindent=0pt
\begin{narrow}[0.4in]}
\def\endtitle{\end{narrow}\newpage\pagenumstyle{arabic}}


\def\beginexercise{\vskip 20truept\parindent=0pt\begin{narrow}[10
truept]}
\def\endexercise{\vskip 10truept\end{narrow}}


\def\eql#1{\eqno\eqnlabel{#1}}
\def\ref{\reference}
\def\peq{\puteqn}
\def\pref{\putref}

\def\mgn{\marginnote}
\def\bex{\begin{exercise}}
\def\eex{\end{exercise}}


\font\open=msbm10 

\font\goth=eufm10  
\font\ssb=cmss10

\def\StretchRtArr#1{{\count255=0\loop\relbar\joinrel\advance\count255 by1
\ifnum\count255<#1\repeat\rightarrow}}
\def\StretchLtArr#1{\,{\leftarrow\!\!\count255=0\loop\relbar
\joinrel\advance\count255 by1\ifnum\count255<#1\repeat}}

\def\StretchLRtArr#1{\,{\leftarrow\!\!\count255=0\loop\relbar\joinrel\advance
\count255 by1\ifnum\count255<#1\repeat\rightarrow\,\,}}

\def\mbox#1{{\leavevmode\hbox{#1}}}

\def\hspace#1{{\phantom{\mbox#1}}}
\def\oR{\mbox{\open\char82}}

\def\oZ{\mbox{\open\char90}}
\def\oC{\mbox{\open\char67}}

\def\gx{\mbox{{\goth\char120}}}

\def\ssa{\mbox{{\ssb\char97}}}

\def\dfone{\mbox{{\ssb\char49}}}
\def\sstwo{\mbox{{\ssb\char50}}}
\def\ssthree{\mbox{{\ssb\char51}}}

\def\al{\alpha}
\def\bsig{{\bmit\sigma}} 

\def\ga{\gamma}

\def\Ga{\Gamma}

\def\la{\lambda}

\def\si{\sigma}

\def\th{\theta}

\def\De{\Delta}

\def\caD{{\cal D}}

\def\caY{{\cal Y}}
\def\caS{{\cal S}}


\def\frac#1/#2{\leavevmode\kern.1em
\raise.5ex\hbox{\the\scriptfont0 #1}\kern-.1em/\kern-.15em
\lower.25ex\hbox{\the\scriptfont0 #2}}
\def\sfrac#1/#2{\leavevmode\kern.1em
\raise.5ex\hbox{\the\scriptscriptfont0 #1}\kern-.1em/\kern-.15em
\lower.25ex\hbox{\the\scriptscriptfont0 #2}}

\def\gtorder{\mathrel{\raise.3ex\hbox{$>$}\mkern-14mu
             \lower0.6ex\hbox{$\sim$}}}
\def\ltorder{\mathrel{\raise.3ex\hbox{$<$}\mkern-14mu
             \lower0.6ex\hbox{$\sim$}}}

\def\semidirprod{\rlap{\ss C}\raise1pt\hbox{$\mkern.75mu\times$}}
\def\for{\lower6pt\hbox{$\Big|$}}
\def\fish{\kern-.25em{\phantom{abcde}\over \phantom{abcde}}\kern-.25em}


\def\boxit#1{\vbox{\hrule\hbox{\vrule\kern3pt
        \vbox{\kern3pt#1\kern3pt}\kern3pt\vrule}\hrule}}
\def\dalemb#1#2{{\vbox{\hrule height .#2pt
        \hbox{\vrule width.#2pt height#1pt \kern#1pt \vrule
                width.#2pt} \hrule height.#2pt}}}
\def\square{\mathord{\dalemb{5.9}{6}\hbox{\hskip1pt}}}

\def\frac#1#2{{{#1}\over{#2}}}

\def\noin{\noindent}


\def\cosec{{\rm cosec\,}}

\def\eg{{\it e.g.}}
\def\ie{{\it i.e. }}
\def\cf{{\it cf }}
\def\pa{\partial}



\def\Threej#1#2#3#4#5#6{\biggl({#1\atop#4}{#2\atop#5}{#3\atop#6}\biggr)}

\def\3j#1#2#3#4#5#6{\left\lgroup\matrix{#1&#2&#3\cr#4&#5&#6\cr}
\right\rgroup}

\def\m?{\mgn{?}}

\def\pa{\partial}

\def\beq{\begin{eqnarray}}
\def\eeq{\end{eqnarray}}


\def\aop#1#2#3{{\it Ann. Phys.} {\bf {#1}} ({#2}) #3}

\def\cmp#1#2#3{{\it Comm. Math. Phys.} {\bf {#1}} ({#2}) #3}
\def\cqg#1#2#3{{\it Class. Quant. Grav.} {\bf {#1}} ({#2}) #3}

\def\ijmp#1#2#3{{\it Int. J. Mod. Phys.} {\bf {#1}} ({#2}) #3}

\def\jmp#1#2#3{{\it J. Math. Phys.} {\bf {#1}} ({#2}) #3}
\def\jpa#1#2#3{{\it J. Phys.} {\bf A{#1}} ({#2}) #3}
\def\jpc#1#2#3{{\it J. Phys.} {\bf C{#1}} ({#2}) #3}
\def\lnm#1#2#3{{\it Lect. Notes Math.} {\bf {#1}} ({#2}) #3}

\def\np#1#2#3{{\it Nucl. Phys.} {\bf B{#1}} ({#2}) #3}
\def\npa#1#2#3{{\it Nucl. Phys.} {\bf A{#1}} ({#2}) #3}
\def\pl#1#2#3{{\it Phys. Lett.} {\bf {#1}} ({#2}) #3}

\def\prp#1#2#3{{\it Phys. Rep.} {\bf {#1}} ({#2}) #3}
\def\pr#1#2#3{{\it Phys. Rev.} {\bf {#1}} ({#2}) #3}
\def\prA#1#2#3{{\it Phys. Rev.} {\bf A{#1}} ({#2}) #3}

\def\prD#1#2#3{{\it Phys. Rev.} {\bf D{#1}} ({#2}) #3}
\def\prl#1#2#3{{\it Phys. Rev. Lett.} {\bf #1} ({#2}) #3}

\def\rmp#1#2#3{{\it Rev. Mod. Phys.} {\bf {#1}} ({#2}) #3}

\def\zfp#1#2#3{{\it Z. f. Phys.} {\bf {#1}} ({#2}) #3}

\def\cras#1#2#3{{\it Comptes Rend. Acad. Sci. (Paris)} {\bf{#1}} (#2) #3}
\def\prs#1#2#3{{\it Proc. Roy. Soc.} {\bf A{#1}} ({#2}) #3}
\def\pcps#1#2#3{{\it Proc. Camb. Phil. Soc.} {\bf{#1}} ({#2}) #3}
\def\mpcps#1#2#3{{\it Math. Proc. Camb. Phil. Soc.} {\bf{#1}} ({#2}) #3}

\def\amsh#1#2#3{{\it Abh. Math. Sem. Ham.} {\bf {#1}} ({#2}) #3}
\def\am#1#2#3{{\it Acta Mathematica} {\bf {#1}} ({#2}) #3}
\def\aim#1#2#3{{\it Adv. in Math.} {\bf {#1}} ({#2}) #3}
\def\ajm#1#2#3{{\it Am. J. Math.} {\bf {#1}} ({#2}) #3}
\def\amm#1#2#3{{\it Am. Math. Mon.} {\bf {#1}} ({#2}) #3}

\def\aom#1#2#3{{\it Ann. of Math.} {\bf {#1}} ({#2}) #3}
\def\cjm#1#2#3{{\it Can. J. Math.} {\bf {#1}} ({#2}) #3}
\def\bams#1#2#3{{\it Bull.Am.Math.Soc.} {\bf {#1}} ({#2}) #3}

\def\cmh#1#2#3{{\it Comm. Math. Helv.} {\bf {#1}} ({#2}) #3}

\def\dmj#1#2#3{{\it Duke Math. J.} {\bf {#1}} ({#2}) #3}
\def\invm#1#2#3{{\it Invent. Math.} {\bf {#1}} ({#2}) #3}

\def\jdg#1#2#3{{\it J. Diff. Geom.} {\bf {#1}} ({#2}) #3}

\def\joa#1#2#3{{\it J. of Algebra} {\bf {#1}} ({#2}) #3}
\def\jram#1#2#3{{\it J. f. reine u. Angew. Math.} {\bf {#1}} ({#2}) #3}
\def\jims#1#2#3{{\it J. Indian. Math. Soc.} {\bf {#1}} ({#2}) #3}
\def\jlms#1#2#3{{\it J. Lond. Math. Soc.} {\bf {#1}} ({#2}) #3}
\def\jmpa#1#2#3{{\it J. Math. Pures. Appl.} {\bf {#1}} ({#2}) #3}
\def\ma#1#2#3{{\it Math. Ann.} {\bf {#1}} ({#2}) #3}

\def\mz#1#2#3{{\it Math. Zeit.} {\bf {#1}} ({#2}) #3}
\def\ojm#1#2#3{{\it Osaka J.Math.} {\bf {#1}} ({#2}) #3}

\def\pems#1#2#3{{\it Proc. Edin. Math. Soc.} {\bf {#1}} ({#2}) #3}

\def\plb#1#2#3{{\it Phys. Letts.} {\bf {B#1}} ({#2}) #3}
\def\pla#1#2#3{{\it Phys. Letts.} {\bf {A#1}} ({#2}) #3}
\def\plms#1#2#3{{\it Proc. Lond. Math. Soc.} {\bf {#1}} ({#2}) #3}
\def\pgma#1#2#3{{\it Proc. Glasgow Math. Ass.} {\bf {#1}} ({#2}) #3}
\def\qjm#1#2#3{{\it Quart. J. Math.} {\bf {#1}} ({#2}) #3}
\def\qjpam#1#2#3{{\it Quart. J. Pure and Appl. Math.} {\bf {#1}} ({#2}) #3}

\def\rmjm#1#2#3{{\it Rocky Mountain J. Math.} {\bf {#1}} ({#2}) #3}

\def\tams#1#2#3{{\it Trans.Am.Math.Soc.} {\bf {#1}} ({#2}) #3}

\begin{title}
\vglue 0.5truein
\vskip15truept
\centertext {\Bigfonts \bf Harmonics on the factored} \vskip7truept
\vskip10truept\centertext{\Bigfonts \bf  three-sphere and the Hopf map}
 \vskip 20truept
\centertext{J.S.Dowker\footnote{dowker@man.ac.uk}} \vskip 7truept
\centertext{\it Theory Group,} \centertext{\it School of Physics and
Astronomy,} \centertext{\it The University of Manchester,}
\centertext{\it Manchester, England} \vskip 7truept \centertext{}

\vskip 7truept

\vskip40truept
\begin{narrow}
Laplacian eigenmodes on homogeneous Clifford--Klein factors of the
three--sphere are obtained as pullbacks of harmonics on the orbifolded
two--sphere using the Hopf map. A method of obtaining these polyhedral,
or crystal, harmonics using binary invariants is presented which has
computational advantages over those based on projection techniques, or
those using invariants constructed in terms of Cartesian coordinates. In
addition, modes transforming according to the irreps of the deck group
are found in easy fashion using the covariants already conveniently
calculated by Desmier and Sharp and by Bellon. Agreement is found with
existing results.
\end{narrow}
\vskip 5truept
\vskip 60truept
\vfil
\end{title}
\pagenum=0
\newpage

\section{\bf 1. Introduction.}

The Hopf map gives the structure of the three--sphere as a fibering, with
a two sphere as base and a circle as fibre. The projection S$^3\to$ S$^2$
can be expressed in terms of the Cartesian coordinates on the
corresponding embedding $\oR^4$ and $\oR^3$ by  the non--linear mapping
(Hopf [\pref{Hopf}]),
  $$\eqalign{
  y^1_{\pm}&=2(x^2x^0\mp x^1x^3)\cr
  y^2_{\pm}&=2(x^2x^3\pm x^1x^0)\cr
  y^3_{\pm}&=(x^3)^2+(x^0)^2-(x^1)^2-(x^2)^2\,.
  }
  \eql{hopfmap}
  $$

The pullback of a harmonic polynomial in a set of $y$s is a harmonic
polynomial in the $x$s. In this paper I wish to expand on this fact,
which is useful because it relates eigenfunctions on S$^2$ and S$^3$.
These two facts are really the same.

Kibler {\it et al}, [\pref{KNR}], have used the Hopf mapping to relate
modes. Hage Hassan and Kibler, [\pref{HHandK}], have elaborated on this
aspect using the Fock--Bargmann--Schwinger approach to angular momentum
theory. Some of the material I give in the next section will be found in
these works, and elsewhere.

A fibre bundle treatment is given by Boiteaux, [\pref{Boiteaux}], (see
also Gilkey {\it et al}, [\pref{GLP}]) but my discussion will be more
simple minded.

In later sections I extend the discussion to include homogeneous factors
of S$^3$ and will give a practical means of obtaining the modes. In this
regard, the Hopf approach has also been expounded by Weeks,
[{\pref{Weeks2}], and Lachi\`eze-Rey and Weeks, [\pref{LandW}], in an
astrophysical context and I refer to these works for references to some
other mode calculations. I should also mention the interesting work of
Bellon, [\pref{Bellon2}], [\pref{Bellon}].

An important step on the road to modes on the factored three--sphere is
the construction of modes on the orbifolded two--sphere. These polyhedral
harmonics, also known as crystal, or lattice, harmonics, have been
investigated for many years; early papers being by  Elert,
[\pref{Elert}], and Bethe, [\pref{Bethe}]. In particular, more recently,
icosahedral harmonics have found use in the study of viruses, fullerenes
and quasi--crystals. The method presented here gives a relatively simple
way of finding them and information in [\pref{Bellon2}] allows the tensor
icosahedral harmonics to be found too.
\section{\bf 2. Harmonic polynomials and eigenmodes}

Working with polynomials has its attractions, but the pullback result is
most easily appreciated in angular coordinates. The Euler angle
expression for the $x^i$, going back at least to Klein and Sommerfeld, is
(I repeat some standard relations),\footnote{ My Euler angles correspond
to the `$z$-$y$-$z$' rotation convention, adopted by [\pref{BandL}],
[\pref{BandS}]. Note that Vilenkin, [\pref{Vilenkin}], uses the other
popular choice, $z$-$x$-$z$.}
  $$\eqalign{
  x^1&=-R\sin \th/2\,\sin (\psi-\phi)/2\cr
  x^2&=R\sin \th/2\,\cos (\psi-\phi)/2\cr
  x^3&=R\cos \th/2\,\sin (\psi+\phi)/2\cr
  x^0&=R\cos \th/2\,\cos (\psi+\phi)/2\,,\cr
  }
  \eql{euler}
  $$
whence, from (\peq{hopfmap}), the $\psi$--independent, and
$\phi$--independent, combinations,
  $$\eqalign{
  y^1_+&=r\sin\th\,\cos\phi\,,\quad y^1_-=r\sin\th\,\cos\psi\cr
  y^2_+&=r\sin\th\,\sin\phi\,,\quad y^2_-=r\sin\th\,\sin\psi\cr
  y^3_+&=r\cos\th\,,\quad\quad\quad\, y^3_-=r\cos\th \cr
  }
  \eql{polar}
  $$
with $r=R^2$, implying the map S$^3\to$ S$^2$. In fact, S$^2$ lifts to a
quartic surface in $\oR^4$ which is the product of two (identical)
three-- spheres.

I have chosen the factors so that the SU(2) element, $U$, agrees in form
with that in [\pref{BandL}], equn.(2.21), {\it viz},
  $$
  U=\caD^{1/2}=x^0{\bf1}-i{\bf x}.\bsig
  \eql{U2}
  $$
in terms of Pauli matrices.

Each set of $y$s in (\peq{polar}) produces a system of spherical polar
coordinates on a corresponding $\oR^3$. I will use just the $y_+$, and
denote these by $y^i$. A complete set of Laplacian eigenfuctions on S$^2$
is provided by the spherical harmonics, $C_l^m(\th,\phi)$. The relation
between these and the SU(2) rep matrices,
$\caD^{j\,\,n}_m(\th,\phi,\psi)$, is (\eg\ Brink and Satchler,
[\pref{BandS}], Vilenkin, [\pref{Vilenkin}], Talman, [\pref{Talman}]),
  $$
   C_l^m(\th,\phi)=(-1)^m\,{\caD^{\,l\,\,\,\,0}_{-m}}(\th,\phi,\psi)\,,\quad
   l\in\oZ\,,
   \eql{spharm}
  $$
the right--hand side being independent of the fibre angle, $\psi$. Since
the $\caD$s are Laplacian eigenfunctions on S$^3\sim$ SU(2) the result
follows. It is standard (\eg\ Hill, [\pref{Hill}], [\pref{BandL}]), that
the solid quantity, $R^{2l}\,\caD^l$, is a harmonic polynomial in the
$x^i$ of degree $2l$ and the corresponding S$^3$ eigenvalues equal
$4l(l+1)$ agreeing with the S$^2$ eigenvalues from (\peq{spharm}) taking
into account the scaling relation between the relevant Laplacians (see
below).

This can be made more explicit by introducing the relevant differential
operators such as the right--generators of the SO(4) action on S$^3$,
  $$\eqalign{
   Y_1&={i\over\sqrt2}\,e^{i\psi}\bigg(i{\pa\over\pa\th}
   +{1\over\sin\th}{\pa\over\pa\phi}
   -\cot\th{\pa\over\pa\psi}
   \bigg)\cr
   Y_{-1}&=Y_1^*\cr
   Y_0&=i{\pa\over\pa\psi}\,,
   }
   \eql{right}
   $$
the first two of which allow the right index in (\peq{spharm}) to be
raised and lowered by the standard action,
   $$
  -iY_m\caD^j(g)=\caD^j(g)\,J_m\,,
  \eql{righta}
  $$
where $J_m$ are the spin--$j$ angular momentum matrices.

The Laplacian, $\De_3$, on the unit three--sphere is the Casimir
operator,
  $$\eqalign{
  \De_3 &=4{\bf Y}^2=4\big(Y_1Y_{-1}+Y_{-1}Y_1-Y_0^2\big)\cr
  &=4\bigg(\pa^2_\th+\cosec^2\th\big(\pa_\phi^2+\pa_\psi^2\big)
  -2\cos\th\,\cosec^2\,\th\,\pa_\phi\,\pa_\psi+\cot\th\,\pa_\th\bigg)\,.
  }
  \eql{3lap}
  $$

This corresponds to the metric,\footnote{ The numerical factor can be
checked by computing the volume of the three--sphere using the
conventional ranges
$0\le\th\le\pi\,,0\le\phi\le2\pi\,,-2\pi\le\psi\le2\pi$ (but see
[\pref{BandL}], [\pref{JandDeV}]), and the metric determinant $\sqrt
g=\sin\th/8$. The volume is $2\pi^2a^3$ for radius $a$. \cf\
[\pref{GLP}].}
  $$\eqalign{
  ds^2&={1\over4}\big(d\th^2+d\phi^2+d\psi^2+2\cos\th\,d\phi\,d\psi\big)\cr
  &={1\over4}\big(d\th^2+\sin^2\th\,
  d\phi^2+(d\psi+\cos\th\,d\phi)^2\big)\,,
  }
  $$
which occurs as the spatial part of the metric on $\oR^4$,
  $$
  d\gx^2=dR^2 +R^2\,ds^2\,,
  $$
giving the Dalembertian (see \eg\ Hund, [\pref{Hund}]),
  $$
  \square_4={1\over R^3}\,\pa_R\,\big(R^3\,\pa_R\big)+\De_3\,.
  $$

Now let $\square_4$ act on a pullback, \ie on a $\psi$--independent
function, $f_0(\th,\phi)$, \eg\ $C_l^m(\th,\phi)$. Then from
(\peq{3lap}), and using $r=R^2$ and $Y_0\,f_0=0$,
  $$\eqalign{
  \square_4 f_0&=\bigg({1\over R^3}\,\pa_R\,\big(R^3\,\pa_R\big)
  +{1\over R^2}\,\De_3\bigg)f_0\cr
   &=4r\bigg({1\over r^2}\,\pa_r\,\big(r^2\,\pa_r\big)
   +{1\over r^2}\,\De_2\bigg)f_0\cr
   &=4r\,\square_3\,f_0\,,
   }
   \eql{dals}
  $$
where $\De_2$ is the unit two--sphere Laplacian, in $\oR^3$ with
coordinates $y^\al$,
  $$
  \De_2=\pa_\th^2+\cot\th\,\pa_\th+\cosec^2\th\,\pa_\phi^2\,.
  $$

Equation (\peq{dals}) shows that the Hopf mapping pulls harmonic
functions back to harmonic functions. Most of the above is well known, in
one form or another.

\section{\bf 3. Harmonic projection}

It is relevant to consider, briefly, the relation between harmonic
projections in the light of the Hopf map.

A standard result is that, in an $\oR^d$, a rational integral polynomial
of degree $n$, $f_n(\gx)$, can be decomposed as
  $$
   f_n(\gx)=Y_n(\gx)+\gx^{\,2}\,f_{n-2}(\gx)\,.
   \eql{gaussexp}
  $$
$Y_n(\gx)$ is a harmonic polynomial (solid spherical harmonic). For $d=3$
this was proved by Gauss and his method easily extends to
$d$--dimensions.

The polynomial $Y_n$ is the {\it harmonic projection} of $f_n$ and there
is an explicit expression for it, (\eg\ Hobson, [\pref{Hobson1}],
Vilenkin, [\pref{Vilenkin}]),
  $$
   Y_n=H(f_n)=\bigg[1-{\gx^2\,\square_d\over2(2n+d-4)}
   +{\gx^4\,\square_d^2\over2.4(2n+d-4)(2n+d-6)}-
   \ldots \bigg]\,f_n\,,
   \eql{harmproj}
  $$
obtained from the iteration of (\peq{gaussexp}) and applications of
$\square_d$ (\cf\ Clebsch, [\pref{Clebsch}], for $d=3$).

I apply this formula for $d=3$ and $d=4$. In the former case, to accord
with my previous notation, I use $y^\al$ as Cartesian coordinates.  I
also write ${\bf y.y}=r^2$ and, for $d=4$, set $\gx^2=R^2$.

For $d=3$
   $$
   f_n({\bf y})=H\big(f_n({\bf y})\big)+r^{\,2}\,f_{n-2}({\bf y})\,,
   \eql{gaussexp3}
  $$
and
  $$
   H(f_n)=\bigg[1-{r^2\,\square_3\over2(2n-1)}
   +{r^4\,\square_3^2\over2.4(2n-1)(2n-3)}-
   \ldots \bigg]\,f_n\,,
   \eql{harmproj3}
  $$
while for $d=4$
   $$
   F_n(\gx)=H\big(F_n(\gx)\big)+R^{\,2}\,F_{n-2}(\gx)\,,
   \eql{gaussexp4}
  $$
and
  $$
   H(F_n)=\bigg[1-{R^2\,\square_4\over2.2n}
   +{R^4\,\square_4^2\over2.4.2n(2n-2)}-
   \ldots \bigg]\,F_n\,,
   \eql{harmproj4}
  $$

In the special case of the Hopf mapping, harmonic projection commutes
with lifting. If $f_n$ lifts to $F_{2n}$, then $H(f_n)$ lifts to
$H(F_{2n})$ and, if $f_{n-2}$ lifts to $F_{2n-4}$, one has
$F_{2n-2}=R^2F_{2n-4}$ by comparing (\peq{gaussexp3}) and
(\peq{gaussexp4}). Therefore, in this instance, (\peq{gaussexp4}) becomes
  $$
   F_n(\gx)=H\big(F_n(\gx)\big)+R^{\,4}\,F_{n-4}(\gx)\,,
   \eql{gaussexp5}
  $$
and also one sees that (\peq{harmproj3}) lifts to (\peq{harmproj4}). A
direct proof, using the relation between Laplacians, (\peq{dals}), is not
obvious.

The extension to higher dimensions, $d$, of Thomson and Tait's approach
to spherical harmonics is straightforward on noting that $1/\gx^{d-2}$ is
harmonic, $\gx^2\ne0$, with $\gx^2=(x^1)^2+(x^2)^2+\ldots+(x^d)^2$. This
is an ancient fact,\footnote{ To include the point $r=0$ formally, it
would be best to adopt a distributional, Green function approach. See
\eg\ Rowe, [\pref{Rowe}].} and another expression for the harmonic
projection is, [\pref{Hobson1}],
   $$
   H(f_n)=(-1)^n{1\over(d-2)d(d+2)\ldots(d+2n-4)}\gx^{2n+d-2}
   \,f_n(\nabla){1\over\gx^{d-2}}\,.
   \eql{hp}
   $$

In particular, in three dimensions, Maxwell's multipole expression of a
general solid harmonic is,
  $$
  Y_n({\bf y}, {\bf p}_{(k)})=C\gx^{2n+1}\prod_{k=1}^n\big({\bf p}_{(k)}{\bf
  .\nabla}\big)\,{1\over r}\,,
  $$
which depends on the $n$ $3$--vectors, ${\bf p}_{(k)}$, $k=1,\ldots,n$.

\section{\bf 4. Spherical factors}

As described in section 2, lifting the modes from S$^2$ produces only the
integer spin modes on S$^3$. A complete set would also include the
half--odd integer $\caD$s. These can be eliminated by dividing S$^3$ by a
$\oZ_2$ antipodal action (to give the projective three--sphere) and
requiring periodicity. (Relatedly, the Hopf map is unchanged under
parity, $\gx\to-\gx$.) Alternatively, the integer spin modes could be
eliminated by choosing anti--periodicity, as allowed by the topology,
[\pref{Schulman}], [\pref{Dow}], [\pref{DandB}].

The total set of integer--spin modes, $\caD^{l\,\,m'}_m$, is obtained, as
mentioned, by acting with the raising and lowering right operators on the
pullbacks of the S$^2$ modes, (\peq{spharm}), in familiar angular
momentum fashion. Weeks, [\pref{Weeks2}], see also [\pref{LandW}],
describes this process using the polynomial approach and refers to the
index $m'$ as the `twist'. A more conventional name is {\it weight},
coming from invariant theory, via Lie--group theory. It is measured by
the vertical operator, $Y_0$.

Since the answer is already known, there is no especial calculational
merit in finding the modes on S$^3/\oZ_2$ in this way but it does give
them an $\oR^3$, SO(3), geometrical character, arising originally from
the isomorphisms SO(4)$\sim$ SU(2)$\times$ SU(2)$/\oZ_2$ and
SO(3)$\sim$SU(2)/$\oZ_2$ .

It is a particular example of the relation between the spectral problems
on the free action S$^3/\Ga'$ and the orbifolded S$^2/\Ga$ where $\Ga'$
is the double of $\Ga$, noting that $\Ga'=\oZ_2$ when $\Ga={\bf 1}$. The
eigenvalue aspects of this relation have already been expounded and used
in [\pref{Dow11}] and now the geometrical underpinnings are more
apparent, \cf\ [\pref{Weeks2}], [\pref{LandW}], [\pref{Bellon}].

It is only possible to find the modes in this pullback way for a general
(say left) symmetry action, S$^3/\Ga'$, if $\Ga'$ contains an antipodal
$\oZ_2$. This applies to even lens spaces, and when $\Ga'=O',\,T',\,Y'$.

The modes on S$^3/\Ga'$ can be obtained by symmetry adaptation. I define,
to begin with, the left group average, or projection,
  $$\eqalign{
  \phi^{l\,\,\,0}_m(g)&\equiv\bigg[{2l+1\over2\pi^2 a^3|\Ga'|}
  \bigg]^{1/2} \sum_{\ga'\in\,\Ga'} \caD^{l\,\,\,0}_m(\ga'g)\cr
  &=C\, \sum_{\ga'\in\,\Ga'}
  \caD^{l\,\,\,n}_m(\ga')\,\caD^{l\,\,\,0}_n(\th,\phi,\psi)\cr
  &=2C\, \sum_{\ga\in\,\Ga}
  \caD^{l\,\,\,n}_m(\ga)\,\caD^{l\,\,\,0}_n(\th,\phi,\psi)\cr
  &=\bigg[{2l+1\over2\pi^2 a^3|\Ga|}
  \bigg]^{1/2}\, \sum_{\ga\in\,\Ga}
  \caD^{l\,\,\,-n}_m(\ga)\,C_l^n(\th,\phi)\,.\cr
  }
  \eql{persum1}
  $$
The last line is now the preliminary symmetry adaptation of the modes on
the Hopf $(\th,\phi)$--sphere base.\footnote{ I make $\Ga'$ act on the
left so that the coordinates on the sphere are the conventional $\th$ and
$\phi$. The coordinate ranges and boundary identifications are detailed
by Jonker and De Vries, [\pref{JandDeV}], and exhibit the twisted product
structure of S$^3$.} I do not pursue this approach to compute these modes
but will present a better method in the next section.

In the derivation of the relation (\peq{persum1}) between the two
projections, I have used the fact that $\Ga=\Ga'/\oZ_2$ where $\oZ_2$ is
an antipodal action, the non--trivial element of which corresponds to a
rotation through $2\pi$ and is the `central' element, introduced by Bethe
to generate the binary group from the pure rotation one. For $l$
integral, it is equivalent to the identity. Hence the factor of two.

As before, the complete set of (preliminary) modes on S$^3/\Ga'$ is
obtained by acting on the right with the raising and lowering operators.
For a particular eigenvalue (depending on only the label $l$), the
degeneracy is the product of the right degeneracy, which is the range of
the right index, \ie $(2l+1)$, and the left degeneracy, $d(l;\Ga)$,
evaluated by cutting down the overcomplete set of the $\phi^{l\,\,\,0}_m$
modes to the minimal number, usually by diagonalisation. The results were
given in [\pref{Dow}] and derived using the results of Polya and Meyer,
[\pref{Meyer}], who employed Molien's theorem rather than constructing
the group average directly.

Most easily, the (left) degeneracies can be obtained from those on cyclic
groups by making use of the cyclic decomposition of a spectral quantity
$\caS$,
     $$
   \caS(\Ga)={1\over|\Ga|}
   \bigg( \sum_q q\,n_q\, \caS(\oZ_q)-\big(\sum_q
   n_q-1\big)\caS(\oZ_1)\bigg)\,,
   \eql{cycdec2}
  $$
where the group $\Ga$ has $n_q$ axes of order $q$. For those groups
satisfying the orbit--stabiliser relation, $|\Ga|=2qn_q$,
  $$
   \caS(\Ga)={1\over2}
   \bigg( \sum_q\caS(\oZ_q)- \caS(\oZ_1)\bigg)\,.
   \eql{cycdec}
  $$

Relation (\peq{cycdec2}) is given by Meyer, [\pref{Meyer}], and proved in
[\pref{ChandD}]. It is the simplest way of deriving the degeneracies, and
also their generating functions, which satisfy (\peq{cycdec}) and are
given in [\pref{Meyer}] (see also Laporte, [\pref{Laporte}]). For
example, the cyclic ($\oZ_q$) and octahedral ($O$) groups, the generating
functions, defined by
  $$
  g(\si;\Ga)\equiv\sum_{l=0}^\infty d(l;\Ga)\,\si^l\,,
  $$
equal
  $$
  g(\si;\oZ_q)={1\over1-\si}{1+\si^q\over1-\si^q}\,,\quad
  g(\si;O)={1+\si^9\over(1-\si^4)(1-\si^6)}\,.
  \eql{genfun}
  $$

The construction of the modes themselves is not so straightforward.

\section{\bf 5. Scalar polyhedral modes}
The easiest part of determining the independent set of symmetry adapted
modes on S$^3/\Ga'$ is the right raising and lowering, which is routine.
The difficulty lies in the construction of the symmetry adapted spherical
harmonics on S$^2/\Ga$, a topic of great interest to physicists, chemists
and biologists over a long period. A lot of expository and technical
material exists which it is impossible to survey. There are various
approaches. In some, the group average is performed element by element.
In others, it is bypassed, or camouflaged.

I will spend a little time developing a method of calculating the
symmetry adapted modes after which the pullback modes on S$^3/\Ga'$ can
be considered to be known by the process outlined earlier.

Perhaps the method that is most readily automated is one described first
by Hodgkinson, [\pref{Hodgkinson}], and based on the expression
(\peq{harmproj3}). The same technique was developed later, independently
by \eg\ [\pref{PSW}], [\pref{LandH}].

Hodgkinson presented a complete process, if somewhat sketchily,\footnote{
Hodgkinson considers only those modes even under reflections in the
polyhedral symmetry planes. He also does not give any specific
computations. } and complained at the `very heavy labour' involved in the
harmonic projection last step, (\peq{harmproj3}). Nowadays this can be
alleviated using symbolic manipulation, and the algorithm has been
detailed, independently, with this in mind by Ronveaux and Saint--Aubin,
[\pref{RandSA}].

The method depends on the existence of an invariant polynomial integrity
basis whose evaluation is classic, going back, in this setting, to Klein
and with much subsequent work.

The process yields a basis for invariant harmonic polynomials in the
Cartesian coordinates of an embedding $\oR^3$. In the present notation,
these coordinates are the $y^1,y^2,y^3$ of (\peq{polar}). These
polynomials can then be converted to polynomials on S$^3/\Ga'$ via the
relation (\peq{hopfmap}) and the right raising and lowering operators
applied to complete the computation of a full basis, if it is desired to
be so explicit. However, the results would not be illuminating. For large
angular momentum, Cartesian coordinates become unwieldy. Better for the
purpose if the harmonic basis on S$^2/\Ga$ could be expanded in
(standard) spherical harmonics, $C_l^m$ (for fixed $l$), for then, using
the relation (\peq{spharm}) with the irrep matrices, the right raising
and lowering is immediate and consists simply of replacing the right $0$
index by $m'$, running from $l$ to $-l$. This is because left and right
are completely independent for homogeneous factorings.

I will now elaborate on the computational method outlined by Hodgkinson
based on the form (\peq{hp}) rather than on (\peq{harmproj3}).

The tesseral harmonics can be defined in the Thomson and Tait,
[\pref{TandT}], way, (\eg\ Maxwell, [\pref{Maxwell}], Hobson,
[\pref{Hobson}], H\"onl, [\pref{Honl}]),
  $$\eqalign{
   \caY_l^m\equiv P_l^m(\cos\th)\,e^{im\phi}&=N_{lm}\,r^{l+1}
   \,(-1)^m\,{\pa^l\over\pa y_1^m\,\pa
   y_0^{l-m}}{1\over r}\cr
   \caY_l^{-m}\equiv P_l^m(\cos\th)\,e^{-im\phi}&=N_{lm}\,r^{l+1}
   \,{\pa^l\over\pa y_{-1}^m\,\pa
   y_0^{l-m}}{1\over r}\cr
   }
   \eql{leg}
  $$
for $l\ge m\ge0$, $r^2={\bf y.y}$, and
  $$
  N_{lm}=(-i)^l{(2\sqrt2)^m\over{(l-m)}!}\,.
   $$

Instead of the Cartesian components, I use spherical ones, defined by
  $$
  y_1\equiv -{y^1-iy^2\over i\sqrt2}\,,\quad y_{-1}\equiv {y^1+iy^2\over
  i\sqrt2}\,,\quad y_0=-iy^3\,,
  $$
for the `standard' components, and
  $$
  y^1=y_{-1}\,,\quad y^{-1}=y_1\,,\quad y^0=-y_0\,,
  \eql{cc}
  $$
for the contrastandard ones. Also $y_1^*=y_{-1}$ and $y_0^*=-y_0$.

If the Cartesian polynomials are real then one will need the
combinations, ({\it unnormalised} spherical harmonics),
  $$\eqalign{
  {\caY_c\,}_l^m&\equiv P_l^m(\cos\th)\,\cos m\phi={1\over2}N_{lm}
  \,r^{l+1}\,{\pa^{l-m}\over\pa{y_0^{l-m}}}\bigg(
  {\pa^m\over\pa y_{1}^m}+(-1)^m{\pa^m\over\pa y_{-1}^m}\bigg){1\over
  r}\cr
  {\caY_s\,}_l^m&\equiv P_l^m(\cos\th)\,\sin m\phi={1\over2i}N_{lm}
  \,r^{l+1}\,{\pa^{l-m}\over\pa{y_0^{l-m}}}\bigg(
  {\pa^m\over\pa y_{1}^m}-(-1)^m{\pa^m\over\pa y_{-1}^m}\bigg){1\over
  r}\,.
  }
  \eql{real}
  $$
These are the functions in terms of which the existing expressions for
the polyhedral harmonics are written. They are not, formally, the most
convenient. A neater organisation is given at the end of this section.

In (\peq{leg}), the $P_l^m$ are the usual Legendre polynomials, in terms
of which the usual surface spherical harmonics are, [\pref{BandS}],
  $$
   C_l^m(\th,\phi)=(-1)^m\,\bigg[{(l-m)!\over(l+m)!}\bigg]^{1/2}
   P_l^m(\cos\th)\,e^{im\phi}\,,\quad m\ge0\,,
   \eql{usph}
  $$
and so,
  $$\eqalign{
  2\bigg[{(l-m)!\over(l+m)!}\bigg]^{1/2}{\caY_c\,}_l^m(\th,\phi)
  =(-1)^m\,C_l^m+C_l^{-m}\,,\quad m\ge0\cr
  2i\bigg[{(l-m)!\over(l+m)!}\bigg]^{1/2}{\caY_s\,}_l^m(\th,\phi)
  =(-1)^m\,C_l^m-C_l^{-m}\,,\quad m\ge0
  }
  \eql{real2}
  $$
corresponding to (\peq{real}).

The principle now to be employed, founded on (\peq{hp}), is that a basis
for invariant harmonic polynomials is provided by the action of the set
of independent invariant polynomial operators,
$Q(\pa_{y^1},\pa_{y^2},\pa_{y^3})=Q(\nabla_y)$, on $1/r$, (Poole,
[\pref{Poole}], Meyer, [\pref{Meyer}], and Laporte, [\pref{Laporte}]).
This set is built algebraically from the invariant polynomial integrity
basis.

This technique is no different from the one mentioned above, only that,
in using (\peq{leg}) the harmonic projection, (\peq{hp}), has really
already been performed. It yields the polynomial series for the $P_l^m$,
[\pref{Hobson}]. \S85.

In $Q(\nabla_y)$, the three--vector $\nabla_y$ is effectively null,
  $$
  \De=\nabla_y.\nabla_y=2{\pa^2\over\pa y_1\,\pa y_{-1}}-{\pa^2\over\pa
  y_0^2}=2{\pa^2\over\pa y^{-1}\,\pa y^{1}}-{\pa^2\over\pa
  {y^0}^2}=0\,.
  \eql{null3}
  $$

Hodgkinson and Poole derived the required integrity basis {\it directly}
from the rotational 3--geometry of the polyhedra (the icosahedron in
their case) and {\it then} imposed the null--vector condition
(\peq{null3}). Klein, [\pref{Klein}], p.238, calculates the polynomials
as higher polars of binary forms (found from geometry) to which they
return, up to a factor, on enforcing the null condition, (\peq{null3}),
or, rather, its dual. It is, therefore, logically more satisfactory to
start from these binary forms.\footnote{ A detailed account of the
construction of the integrity bases using 3--vectors is given in Appendix
C of [\pref{JMS}].}

Following H\"onl, [\pref{Honl}], define, in a symbolic way, the binary
pseudo--operators,
  $$
  \la_{1/2}=\bigg({\pa\over\pa\, y^{1}}\bigg)^{1/2}
  =\bigg({\pa\over\pa\, y_{-1}}\bigg)^{1/2}\,,\quad
  \la_{-1/2}=\bigg({\pa\over\pa\, y^{-1}}\bigg)^{1/2}
  =\bigg({\pa\over\pa\, y_{1}}\bigg)^{1/2}\,,
  \eql{pop}
  $$
so that, in accord with (\peq{null3}),
  $$
  {\pa\over\pa y^0}=\pm\sqrt2\,\la_{1/2}\,\la_{-1/2}\,.
  \eql{op}
  $$

With these correspondances, and the upper sign in (\peq{op}),\footnote{
In this case, for consistency because of (\peq{cc}), it is necessary to
make the negative sign correspond to the complex conjugate, $(\pa/\pa
y_0)^*=-\sqrt2\,\la_{1/2}\la_{-1/2}$.} it can easily be checked that $
\la_{1/2}$ and $ \la_{-1/2}$ transform as standard spin--1/2 spinors
under SU(2) (\cf\ the interesting paper by H\"onl, [\pref{Honl}]) and
hence invariant polynomial operators can be obtained from Klein's
invariant binary forms on making the replacements $z_1\to\la_{1/2}$ and
$z_2\to\la_{-1/2}$ for his binary variables. For notational, and
comparison, convenience, I will use $z_1$ and $z_2$ for the {\it
operators}, as well.

Klein's fundamental icosahedral binary invariant ground form is the
 12--ic (spin--6),
  $$
  f=z_1\,z_2\big(z_1^{10}-z_2^{10}\big)+11 z_1^6\,z_2^6
  \eql{eff}
  $$
in canonical form.

This has the associated Hessian (spin--10),
  $$
  H=-\big(z_1^{20}+z_2^{20}\big)+
  228\,(z_1z_2)^5\big(z_1^{10}-z_2^{10}\big)-494\,z_1^{10}z_2^{10}\,.
  \eql{H}
  $$
and the Jacobian transvectant (spin--15),
  $$
  T=\big(z_1^{30}+z_2^{30}\big)+522\,(z_1z_2)^5\big(z_1^{20}-z_2^{20}\big)-
  10005\,(z_1z_2)^{10}\big(z_1^{20}+z_2^{20}\big)\,.
  $$

The tesseral harmonics (\peq{real}) can be rewritten, with the operator
replacements,
  $$\eqalign{
  N_l(l-m)!
  \,{\caY_c\,}_l^m=r^{l+1}(z_1z_2)^{l-m}\bigg(
  z_2^{2m}+(-1)^m z_1^{2m}\bigg){1\over r}\cr
  iN_l(l-m)!
  \,{\caY_s\,}_l^m=r^{l+1}(z_1z_2)^{l-m}\bigg(
  z_2^{2m}-(-1)^m z_1^{2m}\bigg){1\over r}\,,
  }
  \eql{real3}
  $$
where $N_l=i^l2^{l/2}/2$ is an irrelevant \ie\ overall, constant.

As examples, one has  the invariant harmonic forms $r^7f \,{1/r}$,
$r^{11}H\,1/r$ and $r^{16}\,T\,1/r$, given by,
  $$\eqalign{
  r^7\,f{1\over r}&\sim 3960\,{\caY_c\,}_6^0-{\caY_c\,}_6^5\cr
  &\sim\caD^{6\,\,0}_{\,\,0}+\sqrt{7/11}\,
  \big(\caD^{6\,\,0}_{-5}-\caD^{6\,\,0}_{\,\,5}\big)\cr
  r^{11}\,H{1\over r}&\sim{\caY_c\,}_{10}^{10}+228\,\,5!\,
  {\caY_c\,}_{10}^5+247\,\,10!\,{\caY_c\,}_{10}^0\cr
  &\sim\caD^{10\,\,0}_{\,\,0}
  -\sqrt{429}/13\,\big(\caD^{10\,\,0}_{-5}
  -\caD^{10\,\,0}_{\,\,\,5}\big)+\sqrt{46189}/247\,
  \big(\caD^{10\,\,\,0}_{-10}+\caD^{10\,\,0}_{\,\,10}\big)\cr
  r^{16}\,T{1\over r}&\sim {\caY_s\,}_{15}^{15}-522\,\,5!\,{\caY_s\,}_{15}^{10}-
  1000510\,\,10!\,\,{\caY_s\,}_{15}^5\cr
  &\sim
\big(\caD^{15\,\,0}_{-5}+\caD^{15\,\,0}_{\,\,5}\big)-
  87\sqrt{7590}/1000510\,\big(\caD^{15\,\,\,0}_{-10}
  -\caD^{15\,\,0}_{\,10}\big)\cr&\hspace{********}-
 3\sqrt{3338335}/1000510\, \big(\caD^{15\,\,\,0}_{-15}
 +\caD^{15\,\,0}_{\,\,15}\big)\,,
  }
  \eql{fht}
  $$
where the $\sim$ sign indicates equality up to an (irrelevant) overall
constant factor.

The expressions (\peq{fht}) agree with those listed by Zheng and
Doerschuk, [\pref{ZandD}], who obtained them by a complicated method from
first principles using projection. Making use of Klein's work in deriving
his forms, many evaluations can be done by hand.

The modes, (\peq{fht}), pull back to vertical modes  on S$^3/\Ga'$. The
remaining modes are obtained, in accordance with previous remarks, simply
by replacing the right $0$ index on the $\caD$s by $m$, with $-l\le m\le
l$, for each appropriate $l$.

A somewhat similar technique is given by Kramer, [\pref{Kramer}],
although he starts with the highest weight modes on S$^3$,
$\caD^{l\,\,\,l}_{m}$ (in my conventions), and lowers the $l$ index. This
means he does not use the pullback description and works only on the
three--sphere.

The general method for the construction of all polyhedral harmonics is
given by Hodgkinson; see also Laporte [\pref{Laporte}] and Meyer,
[\pref{Meyer}]. The number of independent invariant harmonics is the
number of terms of the form $f^a\,H^b\,T^c$ in the general polynomial,
homogeneous of degree $2l$ as $z_1$ and $z_2$ are equally scaled.
However, because of the necessary syzygy,
 $$
 T^2=1728f^5-H^3\,,
 \eql{icossyz}
 $$
$c$ is either zero (even, Neumann modes) or one (odd, Dirichlet modes),
[\pref{Laporte}]. The number of modes, $d_l$, is encoded in the
icosahedral generating function, [\pref{Laporte}], [\pref{Meyer}],
  $$
   g(\si;Y)={1+\si^{15}\over(1-\si^6)(1-\si^{10})}=\sum_ld(l;Y)\,\si^l\,,
  $$
the numerator, $\si^{15}$, corresponding to the odd modes \footnote{ For
the purely rotational situation, there is no obligation to refer to even
and odd modes. The numerator simply reflects the possible existence of a
single factor of $T$, which is related to the syzygy. Also, the
generating function is being derived here from the integrity basis.
Often, the generating function is derived by other means (\eg\ by cyclic
decomposition) and information about the basis obtained therefrom, \eg\
Sloane, [\pref{Sloane}], [\pref{PSW}].}.

In order to expand the harmonic in any particular case, set
   $$\eqalign{
     r^{l+1}f^a\,H^b\,T^c{1\over r}=
     \sum_{m=0}^{l=l_c}(l-m)!\,A^{c\,l}_m\,{\caY_s\,}_l^m
    }
    \eql{exp}
   $$
with the upper limit $l_c=6a+10b+15c$, $c=0,1$. It can be seen that the
problem reduces to the algebraic one of organising the polynomial $
f^a\,H^b\,T^c$ into a sum of the expressions appearing on the right--hand
side of (\peq{real3}), [\pref{Hodgkinson}].

For this purpose, it is sufficient to set $z_1=1$, so that, from
(\peq{real3}), we seek
   $$\eqalign{
     f^a\,H^b\,T^c\bigg|_{z_1=1}
     &\sim\sum_{m=0}^{l=l_c}A^{c\,l}_m\,\big(z^{l+m}
     +(-1)^{m+c}\,z^{l-m}\big)\,,\cr
     }
     \eql{gent}
   $$
and, as indicated, I am working up to an irrelevant overall constant.

Evaluating the left--hand side using (\peq{eff}) and (\peq{H}), the
coefficients $A^{c\,l}_m$ can be read off yielding the expansion
coefficients in (\peq{exp}). In more detail,
  $$\eqalign{
  (z^{11}-11\, z^6-z)^a\,&(-z^{20}-228\,z^{15}-494\,z^{10}+228\,z^5-1)^b\cr
  &\times(z^{30}-522\,z^{25}-10005\,z^{20}-10005\,z^{10}+522\,z^5+1)^c\cr
  &=\sum_{m=0,5,10,\ldots}^{6a+10b+15c}
  A^{c\,l}_m\,\big(z^{l+m}+(-1)^{m+c}\,z^{l-m}\big)\,.
  }
  \eql{leven}
  $$
Equation (\peq{leven}) for $c=0$ agrees with the general form given by
Poole, [\pref{Poole2}], equn.(11).

Machine algebra allows one to proceed as far as required. For a given $l$
it is only necessary to determine the possible values of $a$, and $b$, an
easy task, and then extract some polynomial coefficients by commonplace
routine. An example is $l=30$ for which $c=0$ and there are two values
for $(a,b)$, namely $(5,0)$ and $(0,3)$. Correspondingly, one obtains two
modes from (\peq{leven}). I find for these,
  $$\eqalign{
  \Psi_{30}^{(5,0)}\sim5!{\caY_c\,}_{30}^{25}-10!\,55\,{\caY_c\,}_{30}^{20}
  +15!1205&\,{\caY_c\,}_{30}^{15}-20!13090\,
  {\caY_c\,}_{30}^{10}\cr
  &+25!\,69585\,{\caY_c\,}_{30}^{5}-30!\,134761\,{\caY_c\,}_{30}^0
  }
  $$
and
  $$\eqalign{
  \Psi_{30}^{(0,3)}\sim{\caY_c\,}_{30}^{30}+&5!\,684{\caY_c\,}_{30}^{25}+10!\,157434\,{\caY_c\,}_{30}^{20}
  +15!\,12527460\,{\caY_c\,}_{30}^{15}\cr
  &+20!77460495\,{\caY_c\,}_{30}^{10}
  +25!\,130689144\,{\caY_c\,}_{30}^{5}-30!\,33211924\,{\caY_c\,}_{30}^0\,.
  }
  $$

The modes listed in [\pref{ZandD}] are linear combinations of these. Thus
  $$\eqalign{
  T_{30,1}&\sim 12251\,\Psi_{30}^{(0,3)}-33211924/11\,\Psi_{30}^{(5,0)}\cr
  T_{30,0}&\sim \,\Psi_{30}^{(0,3)}-45750\,\Psi_{30}^{(5,0)}\,.
  }
  $$

The procedure for the cubic groups is similar. The fundamental ground
form can be taken to be the special sextic,
  $$
  f=z_1z_2(z_1^4-z_2^4)=a_z^6
  \eql{oform}
  $$
(corresponding to the Cartesian, $y^1y^2y^3$) from which the complete
form system can be derived, geometrically,  \eg\ [\pref{Klein2}], or
algebraically by invariant theory, [\pref{Gordan2}].

The Hessian, $H$, and the Jacobian, $T$, of $f$ and $H$ are
  $$\eqalign{
  H=(f,f)^2=(ab)^2a_z^4b_z^4=-{1\over18}\,\big(z_1^8+14\,z_1^4z_2^4+z_2^8\big)\cr
  T=(f,H)=-{1\over108}\,\big(z_1^{12}-33\,z_1^8z_2^4
  -33\,z_1^4z_2^8+z_4^{12}\big)\,,\cr
  }
  $$
where $(f,g)$ is the transvectant of the forms $f$ and $g$, \eg\
[\pref{GandY}].

The form $f$ is an absolute invariant for the tetrahedral group and so a
complete set of absolute invariants is $f$, $H$, and $T$, corresponding
to spins $3$, $4$ and $6$, respectively.

The standard syzygy,
  $$
  T^2=-{1\over108}\,f^4-{1\over2}\,H^3\,,
  \eql{syz}
  $$
means that the general term can be written $f^a\,H^b\,T^c$ where $c=0,1$
and $l=3a+4b+6c$ so that the generating function is,
  $$
   g(\si;T)={1+\si^{6}\over(1-\si^3)(1-\si^{4})}=\sum_ld(l;T)\,\si^l\,.
  $$

By contrast, the octahedral group replaces $f$ by $\pm f$ which means
that the Hessian, $H$, is an absolute invariant, but that $T$ is replaced
by $\pm T$. Therefore a basis set of absolute invariants in this case is
provided by $f^2$, $H$ and $fT$, with spins $6$, $4$ and $9$,
respectively. \eg\ [\pref{Klein}], p.69. Comparing notations with
Laporte, [\pref{Laporte}], $f\sim\phi_3$, $H\sim\phi_4$, $T\sim\Phi_6$,
$fT\sim\Phi_9$, as can be checked algebraically.

The syzygy, (\peq{syz}), is recast as,
  $$
  (fT)^2=-f^2\bigg({1\over108}\,f^4+{1\over2}\,H^3\bigg)\,,
  \eql{osyz}
  $$
so that the term in the general polynomial this time is
$f^{2a'}\,H^b\,(fT)^c$ with $c=0,1$ and $l=6a'+4b+9c$. The octahedral
generating function is therefore,
  $$
   g(\si;O)={1+\si^{9}\over(1-\si^4)(1-\si^{6})}=\sum_ld(l;O)\,\si^l\,,
  $$
given previously.

As a typical example set $l=12$ so that $c=0$ and $(a',b)=(2,0)$ or
$(0,3)$. The general definitions (\peq{exp}) and (\peq{gent}) still apply
with the upper limits $l_c=3a+4b+9c$, where $a=2a'$ and, in place of
(\peq{leven}), one has
  $$\eqalign{
  (z-z^5)^a\,&(1+14z^4+z^8)^b(1-33z^4-33z^8+z^{12})^c\cr
  &=\sum_{m=0,4,8,\ldots}^{3a+4b+9c}
  A^{c\,l}_m\,\big(z^{l+m}+(-1)^{m+c}\,z^{l-m}\big)\,,
  }
  \eql{leveno}
  $$
which readily yields the invariant modes, ($c=0$),
  $$\eqalign{
  \Psi_{12}^{(2,0)}\sim4!\,{\caY_c\,}_{12}^8-8!\,
  {\caY_c\,}_{12}^4+12!\,3\,{\caY_c\,}_{12}^0
   }
  $$
and
  $$\eqalign{
  \Psi_{12}^{(0,3)}\sim{\caY_c\,}_{12}^{12}+
  4!\,42\,{\caY_c\,}_{12}^8+ 8!\,591\,{\caY_c\,}_{12}^4
  +12!\,1414\,{\caY_c\,}_{12}^0\,.
  }
  $$

These quantities agree with the corresponding ones in Table 8 in Altmann
and Bradley, [\pref{AandB}], after a linear combination which
orthogonalises them. I have not checked them against the Cartesian
expressions given by Ronveaux and Saint--Aubin, [\pref{RandSA}],
equn.(34).

Incidentally, it is interesting to note that the fourth transvectant of
the octahedral form $f$ with itself is zero. In fact, this characterises
the form, (\peq{oform}), Klein, [\pref{Klein2}], Cayley,
[\pref{Cayley4}], Gordan, [\pref{Gordan2}]. In angular momentum terms
this means that the vanishing of the Clebsch--Gordan spin--two
combination of two spin--three quantities, $f$,
  $$
  f^l\, f^m\Threej332lmn=0\,,
  $$
implies that $f$ is equivalent to the special octahedral form,
(\peq{oform}).

Similar remarks hold for the icosahedral case and are related to Fuch's
notion of {\it Primformen}.

A further ancillary point concerns the nature of the syzygies. These can
be proved either geometrically, [\pref{Klein}], or algebraically,
[\pref{Gordan2}]. Another approach uses the action under the reflections
of the {\it extended} groups. Taking the icosahedral syzygy,
(\peq{icossyz}) as an example, $T$ is odd under reflection while $H$ and
$f$ are even. Hence $T^2$, being even, can be expressed algebraically in
terms of $f$ and $H$, \eg\ [\pref{Laporte}]. In fact, all the left--hand
sides of the syzygies, (\peq{icossyz}), (\peq{syz}) and (\peq{osyz}) are
squares of odd quantities (Jacobians) and the corresponding modes thus
are counted by the numerator in the generating functions given above.
This is a well known behaviour. The denominator basis elements sometimes
are referred to as `free', and the numerator ones as `constrained' and
occur only once in the construction of the algebraic basis, \eg\ Cummins
and Patera, [\pref{CandP}], Patera {\it et al}, [\pref{PSW}], McLellan,
[\pref{McLellan}].}

Of course the entire scheme is a classic example of the invariants of
finite reflection groups and can be treated from this point of view {\it
ab initio}.

For the cyclic and dihedral groups, the construction of invariant bases
in terms of the Legendre functions proceeds simply by a process of
selection and is given in, \eg, Meyer, [\pref{Meyer}] and, earlier, in
the classic, Pockels, [\pref{Pockels}]. See also Altmann and Bradley,
[\pref{AandB}].

In order to make comparison with existing results easier, I have used the
tesseral harmonics. However it is more elegant to use the usual spherical
harmonics, (\peq{usph}). Then, instead of (\peq{real3}), some
rearrangement yields the neater relation,
  $$\eqalign{
  (-1)^l\,N_l\,
  C^l_m(\th,\phi)=r^{l+1}\,\overrightarrow{Z}^{(l)}_m\,{1\over r}\,,\cr
  }
  \eql{real5}
  $$
with the familiar monomial (or null $l$--spinor),
  $$
  Z^{(l)}_m\equiv
  {z_1^{l+m}z_2^{l-m}\over\big((l-m)!(l+m)!\big)^{1/2}}\,,
  $$
constructed from the two--spinor $\left(\matrix{z_1\cr z_2}\right)$, \cf\
[\pref{Honl}], equn.(28). \vglue2truept However, we have now come full
circle, because (\peq{real5}) can be rewritten
  $$\eqalign{
  {(2l)!\over2^l\,l!}\,C^l_m(\th,\phi)=r^{l+1}
  \,\oC^l_m(-\nabla_y)\,{1\over r}\,,\cr
  }
  \eql{real6}
  $$
where $\oC^l_m(\ssa)$ is the solid spherical harmonic and $\nabla_y$ is a
null vector according to (\peq{null3}). Equation (\peq{real6}) is an
example of Nivens' general theorem, [\pref{Niven}], [\pref{Hobson}],
p.127, and could be taken conveniently as the starting point of the
analysis, rather than the specific (\peq{leg}). It also follows from
(\peq{hp}).

\section{\bf 6. Tensor polyhedral  modes}

The modes found in the last section are invariant, \ie\ they transform as
the identity rep, ${\dfone}$, of $\Ga$. In order to find those that
transform equivariantly according to the other irreps (some are listed in
[\pref{AandB}] for example) I use the results of Patera {\it et al},
[\pref{PSW}], and, especially, of Desmier and Sharp, [\pref{DandSh}] and
of Bellon [\pref{Bellon2}]. I restrict myself to the octahedral and
icosahedral groups.

To start with, the mode (which could be called a twisted scalar mode)
transforming as the one--dimensional rep, ${\dfone'}$, of the octahedral
group has the general form
  $$
  \Psi_l({\dfone'})=r^{l+1}\big(A\,f+B\,T\big){1\over r}
  $$
where $A$ and $B$ are invariants constructed as in the preceding section
from $f^2$ and $H$. For example, for $l=12$, $A=0$ and $B\sim f^2$ and
the mode is easily calculated to be,
  $$
  \Psi_{12}({\dfone'})\sim 10!\,34\,
  {\caY_c}_{12}^2-6!\,35\,{\caY_c}_{12}^6+2!\,{\caY_c}_{12}^{10}\,,
  $$
again in agreement with [\pref{AandB}] Table 8.

The next most complicated modes transform according to the irrep
${\sstwo}$ and have the general structure
   $$
    \Psi_l({\sstwo})=r^{l+1}\overrightarrow{\Psi}_l({\sstwo}){1\over r}
   $$
where, [\pref{DandSh}] (18),
  $$\eqalign{
  \overrightarrow{\Psi}_l&({\sstwo})=\cr
  &A\left(\matrix{\sqrt3\big(z_1^4+6z_1^2z_2^2+z_2^4\big)\cr
  -3(z^2_1-z^2_2)^2}\right)+B\left(\matrix{\sqrt3\big(z_1^8
  +4z_1^6z_2^2-10z_1^4z_2^4+4z_1^2z_2^6+z_2^8\big)\cr
  z_1^8-12z_1^6z_2^2-10z_1^4z_2^4-12z_1^2z_2^6+z_2^8}\right)
  }
  $$
with $A$ and $B$ as before. Choosing again $l=12$, and writing
  $$
  A\sim f^{2a}H^b\,,\quad B\sim f^{2a'}H^{b'}\,,
  $$
the only possibilities are $(a,b)=(1,1)$ and $(a',b')=(0,2)$ giving two
independent modes. I write them out just to show how the method proceeds.

Applying the technique explained in the last section directly to the
$\sstwo$--tensor integrity basis given by Desmier and Sharp yields the
two modes, (I write  $\caY$ for $\caY_c$ for space reasons),
  $$
  \Psi_{12}^1(\sstwo)=\left(\matrix{\sqrt3(-12!78\caY_{12}^0
    -10!14\caY_{12}^2+8!72\caY_{12}^4+6!13\caY_{12}^6
    +4!6\caY_{12}^8+2!\caY_{12}^{10})\cr
    -12!78\caY_{12}^0
    +10!42\caY_{12}^2+8!72\caY_{12}^4-6!39\caY_{12}^6
    +4!6\caY_{12}^8-2!3\caY_{12}^{10}
                       }\right)
  \eql{mode1}
  $$
and
  $$
  \Psi_{12}^2(\sstwo)=\!\!\left(\matrix{\!\!\sqrt3\big(\!\!-\!12!962\caY_{12}^0
    +10!904\caY_{12}^2-8!81\caY_{12}^4+6!116\caY_{12}^6
    +4!18\caY_{12}^8+2!\caY_{12}^{10}\big)\cr
    \big(-12!962\caY_{12}^0
    +10!2712\caY_{12}^2-8!81\caY_{12}^4-6!348\caY_{12}^6
    +4!18\caY_{12}^8\cr\hspace{********}-2!12\caY_{12}^{10}+\caY_{12}^{12}\big)
                       }\right)
   \eql{mode2}
  $$
which are, in appearance, more complicated than those in [\pref{AandB}]
Table 10. The reason is that the basis in $\sstwo$--irrep space used in
[\pref{DandSh}] is different to that in [\pref{AandB}].

In order to convert the modes (\peq{mode1}) and (\peq{mode2}) to Altmann
and Bradley's, firstly multiply them by the SO(2) rotation,
  $$
              R=\left(\matrix{\cos\pi/3&-\sin\pi/3\cr
                              \sin\pi/3&\cos\pi/3}\right)
  $$
to convert the $\sstwo$ basis, and then take linear combinations in order
to get precise agreement. I find
  $$\eqalign{
   \Psi^{AB\,1}_{12}(\sstwo)\sim 37\, R\Psi_{12}^1(\sstwo)-
     3\,R\Psi_{12}^2(\sstwo)\cr
     \Psi^{AB\,2}_{12}(\sstwo)\sim 8 \,R\Psi_{12}^1(\sstwo)
     +R\Psi_{12}^2(\sstwo)\,.
   }
  $$

The same considerations apply to the other irreps of the octahedral (and
the binary octahedral) group but I do not need to give any further
examples. I comment that the $\ssthree$--irrep of O ($\Ga_4$ in
[\pref{DandSh}] \footnote{ There appears to be a misprint in equn.(56) of
[\pref{PSW}]. The irrep for O should be $\Ga_4$.}) requires no rotation
of basis. For example the fundamental $l=1$ mode is,
   $$
   E^{(2)}_{4,6}\sim\left(\matrix {x\cr y\cr z\cr}\right)\,,
   $$
with the notation of [\pref{DandSh}], and so has a Cartesian basis. This
agrees as it stands with the entry in [\pref{AandB}], Table 11. $\Ga_6$
of [\pref{DandSh}] is the spinor, quaternion irrep, $\sstwo_s$.

The general structure of the bases is outlined in Section 4 of
[\pref{DandSh}] and it is clear that any desired mode can be constructed
without too much bother. A basic symbolic manipulator (I used DERIVE) is
all that is required. If one were to pursue this extensively, it would be
advisable to rotate the $\sstwo$--octahedral expressions given in
[\pref{DandSh}] by $R$ at the start.

The generating functions for the icosahedral group are given in Table VII
of [\pref{DandSh}] but not the corresponding reps. However, Bellon,
[\pref{Bellon2}], gives a list of those based on the fundamental
quaternion irrep, $\sstwo_s$, which can be used in exactly the same way
as above to give the corresponding tensor icosahedral harmonics.

A technical point is that the reps in [\pref{Bellon2}] are taken with
respect to a spherical basis and it is easiest to use standard spherical
functions and the basic relation (\peq{real5}).

As an illustration, the $l=1$ and $l=5$ modes for the vector
$\ssthree$--irrep are rapidly found to be proportional to,
  $$
 \displaylines{\eqalign{\left(\matrix{\sqrt2\,C^1_1
 \cr C^1_0\cr \sqrt2\,C^1_{-1}\cr}\right)\quad
{\rm and}\quad\left(\matrix{\sqrt{30}\,C^5_1-\sqrt{70}\,C^5_{-4}\cr
  \sqrt{7}\,C^5_{-5}-6\,C^5_0-\sqrt{7}\,C^5_{5}\cr
  \sqrt{30}\,C^5_{-1}+\sqrt{70}\,C^5_4  \cr
  }\right)\,.
  }}
  $$
respectively. In this form, the lifting to dodecahedron space is easy.

It would need a certain motivation to compute further cases or to check
orthogonality for degenerate modes, although there is no particular
difficulty.

The modes which transform according to the spinor irreps of the binary
group, $\Ga'$, will involve what might be termed polyhedral spinor
harmonics leading to polyhedral spinor hyperharmonics on S$^3/\Ga'$.
These will be dealt with at another time using fractional derivatives.

\section{ \bf 7. Concluding remarks}

I have employed the Hopf map to lift modes on the two--sphere to modes on
the three--sphere, a known procedure, and have divided by a polyhedral
symmetry. I used this to motivate the construction of the required
symmetry adapted modes on the orbifolded two--sphere which are computed
by a binary method and agree with those obtained many years ago by the
more standard, and in my view more involved, technique of projection.
This has been extended to tensor modes under the action of the deck group
and some specific cases have been evaluated using the results of Desmier
and Sharp and of Bellon.

Expressing the symmetry adapted modes in the traditional way as sums of
spherical harmonics allows the complete set of modes on the factored
three--sphere to be found from the pullbacks using raising and lowering
with no extra work.

An aspect that deserves exploration is the significance of the
degeneracies, $d(l;\Ga)$, as the dimensions of the irreps of the symmetry
group of S$^2/\Ga$ which is the centraliser of $\Ga$ in SO(3), or for
S$^3/\Ga'$, the centraliser of the binary group, $\Ga'$, in  SU(2). This
is the `missing label' question.

 \vglue 20truept

 \noin{\bf References.} \vskip5truept
\begin{putreferences}
   \ref{Hund}{Hund,F. \zfp{51}{1928}{1}.}
   \ref{Elert}{Elert,W. \zfp {51}{1928}{8}.}
   \ref{Poole2}{Poole,E.G.C. \qjm{3}{1932}{183}.}
   \ref{Bellon}{Bellon,M.P. {\it On the icosahedron: from two to three
   dimensions}, arXiv:0705.3241.}
   \ref{Bellon2}{Bellon,M.P. \cqg{23}{2006}{7029}.}
   \ref{McLellan}{McLellan,A,G. \jpc{7}{1974}{3326}.}
   \ref{Boiteaux}{Boiteaux, M. \jmp{23}{1982}{1311}.}
   \ref{HHandK}{Hage Hassan,M. and Kibler,M. {\it On Hurwitz
   transformations} in {Le probl\`eme de factorisation de Hurwitz}, Eds.,
   A.Ronveaux and D.Lambert (Fac.Univ.N.D. de la Paix, Namur, 1991),
   pp.1-29.}
   \ref{Weeks2}{Weeks,Jeffrey \cqg{23}{2006}{6971}.}
   \ref{LandW}{Lachi\`eze-Rey,M. and Weeks,Jeffrey, {\it Orbifold construction of
   the modes on the Poincar\'e dodecahedral space}, arXiv:0801.4232.}
   \ref{Cayley4}{Cayley,A. \qjpam{58}{1879}{280}.}
   \ref{JMS}{Jari\'c,M.V., Michel,L. and Sharp,R.T. {\it J.Physique}
   {\bf 45} (1984) 1. }
   \ref{AandB}{Altmann,S.L. and Bradley,C.J.  {\it Phil. Trans. Roy. Soc. Lond.}
   {\bf 255} (1963) 199.}
   \ref{CandP}{Cummins,C.J. and Patera,J. \jmp{29}{1988}{1736}.}
   \ref{Sloane}{Sloane,N.J.A. \amm{84}{1977}{82}.}
   \ref{Gordan2}{Gordan,P. \ma{12}{1877}{147}.}
   \ref{DandSh}{Desmier,P.E. and Sharp,R.T. \jmp{20}{1979}{74}.}
   \ref{Kramer}{Kramer,P., \jpa{38}{2005}{3517}.}
   \ref{Klein2}{Klein, F.\ma{9}{1875}{183}.}
   \ref{Hodgkinson}{Hodgkinson,J. \jlms{10}{1935}{221}.}
   \ref{ZandD}{Zheng,Y. and Doerschuk, P.C. {\it Acta Cryst.} {\bf A52},
   (1996) 221.}
   \ref{Honl}{H\"onl,H. \zfp{89}{1934}{244}.}
   \ref{PSW}{Patera,J., Sharp,R.T. and Winternitz,P. \jmp{19}{1978}{2362}.}
   \ref{LandH}{Lohe,M.A. and Hurst,C.A. \jmp{12}{1971}{1882}.}
   \ref{RandSA}{Ronveaux,A. and Saint-Aubin,Y. \jmp{24}{1983}{1037}.}
   \ref{JandDeV}{Jonker,J.E. and De Vries,E. \npa{105}{1967}{621}.}
   \ref{Rowe}{Rowe, E.G.Peter. \jmp{19}{1978}{1962}.}
   \ref{KNR}{Kibler,M., N\'egadi,T. and Ronveaux,A. {\it The Kustaanheimo-Stiefel
   transformation and certain special functions} \lnm{1171}{1985}{497}.}
   \ref{GLP}{Gilkey,P.B., Leahy,J.V. and Park,J-H, \jpa{29}{1996}{5645}.}
   \ref{Kohler}{K\"ohler,K.: Equivariant Reidemeister torsion on
   symmetric spaces. Math.Ann. {\bf 307}, 57-69 (1997)}
   \ref{Kohler2}{K\"ohler,K.: Equivariant analytic torsion on ${\bf P^nC}$.
   Math.Ann.{\bf 297}, 553-565 (1993) }
   \ref{Kohler3}{K\"ohler,K.: Holomorphic analytic torsion on Hermitian
   symmetric spaces. J.Reine Angew.Math. {\bf 460}, 93-116 (1995)}
   \ref{Zagierzf}{Zagier,D. {\it Zetafunktionen und Quadratische
   K\"orper}, (Springer--Verlag, Berlin, 1981).}
   \ref{Stek}{Stekholschkik,R. {\it Notes on Coxeter transformations and the McKay
   correspondence.} (Springer, Berlin, 2008).}
   \ref{Pesce}{Pesce,H. \cmh {71}{1996}{243}.}
   \ref{Pesce2}{Pesce,H. {\it Contemp. Math} {\bf 173} (1994) 231.}
   \ref{Sutton}{Sutton,C.J. {\it Equivariant isospectrality
   and isospectral deformations on spherical orbifolds}, ArXiv:math/0608567.}
   \ref{Sunada}{Sunada,T. \aom{121}{1985}{169}.}
   \ref{GoandM}{Gornet,R, and McGowan,J. {\it J.Comp. and Math.}
   {\bf 9} (2006) 270.}
   \ref{Suter}{Suter,R. {\it Manusc.Math.} {\bf 122} (2007) 1-21.}
   \ref{Lomont}{Lomont,J.S. {\it Applications of finite groups} (Academic
   Press, New York, 1959).}
   \ref{DandCh2}{Dowker,J.S. and Chang,Peter {\it Analytic torsion on
   spherical factors and tessellations}, arXiv:math.DG/0904.0744 .}
   \ref{Mackey}{Mackey,G. {\it Induced representations}
   (Benjamin, New York, 1968).}
   \ref{Koca}{Koca, {\it Turkish J.Physics}.}
   \ref{Brylinski}{Brylinski, J-L., {\it A correspondence dual to McKay's}
    ArXiv alg-geom/9612003.}
   \ref{Rossmann}{Rossman,W. {\it McKay's correspondence
   and characters of finite subgroups of\break SU(2)} {\it Progress in Math.}
      Birkhauser  (to appear) .}
   \ref{JandL}{James, G. and Liebeck, M. {\it Representations and
   characters of groups} (CUP, Cambridge, 2001).}
   \ref{IandR}{Ito,Y. and Reid,M. {\it The Mckay correspondence for finite
   subgroups of SL(3,C)} Higher dimensional varieties, (Trento 1994),
   221-240, (Berlin, de Gruyter 1996).}
   \ref{BandF}{Bauer,W. and Furutani, K. {\it J.Geom. and Phys.} {\bf
   58} (2008) 64.}
   \ref{Luck}{L\"uck,W. \jdg{37}{1993}{263}.}
   \ref{LandR}{Lott,J. and Rothenberg,M. \jdg{34}{1991}{431}.}
   \ref{DoandKi} {Dowker.J.S. and Kirsten, K. {\it Analysis and Appl.}
   {\bf 3} (2005) 45.}
   \ref{dowtess1}{Dowker,J.S. \cqg{23}{2006}{1}.}
   \ref{dowtess2}{Dowker,J.S. {\it J.Geom. and Phys.} {\bf 57} (2007) 1505.}
   \ref{MHS}{De Melo,T., Hartmann,L. and Spreafico,M. {\it Reidemeister
   Torsion and analytic torsion of discs}, ArXiv:0811.3196.}
   \ref{Vertman}{Vertman, B. {\it Analytic Torsion of a  bounded
   generalized cone}, ArXiv:0808.0449.}
   \ref{WandY} {Weng,L. and You,Y., {\it Int.J. of Math.}{\bf 7} (1996)
   109.}
   \ref{ScandT}{Schwartz, A.S. and Tyupkin,Yu.S. \np{242}{1984}{436}.}
   \ref{AAR}{Andrews, G.E., Askey,R. and Roy,R. {\it Special functions}
  (CUP, Cambridge, 1999).}
   \ref{Tsuchiya}{Tsuchiya, N.: R-torsion and analytic torsion for spherical
   Clifford-Klein manifolds.: J. Fac.Sci., Tokyo Univ. Sect.1 A, Math.
   {\bf 23}, 289-295 (1976).}
   \ref{Tsuchiya2}{Tsuchiya, N. J. Fac.Sci., Tokyo Univ. Sect.1 A, Math.
   {\bf 23}, 289-295 (1976).}
  \ref{Lerch}{Lerch,M. \am{11}{1887}{19}.}
  \ref{Lerch2}{Lerch,M. \am{29}{1905}{333}.}
  \ref{TandS}{Threlfall, W. and Seifert, H. \ma{104}{1930}{1}.}
  \ref{RandS}{Ray, D.B., and Singer, I. \aim{7}{1971}{145}.}
  \ref{RandS2}{Ray, D.B., and Singer, I. {\it Proc.Symp.Pure Math.}
  {\bf 23} (1973) 167.}
  \ref{Jensen}{Jensen,J.L.W.V. \aom{17}{1915-1916}{124}.}
  \ref{Rosenberg}{Rosenberg, S. {\it The Laplacian on a Riemannian Manifold}
  (CUP, Cambridge, 1997).}
  \ref{Nando2}{Nash, C. and O'Connor, D-J. {\it Int.J.Mod.Phys.}
  {\bf A10} (1995) 1779.}
  \ref{Fock}{Fock,V. \zfp{98}{1935}{145}.}
  \ref{Levy}{Levy,M. \prs {204}{1950}{145}.}
  \ref{Schwinger2}{Schwinger,J. \jmp{5}{1964}{1606}.}
  \ref{Muller}{M\"uller, \lnm{}{}{}.}
  \ref{VMK}{Varshalovich.}
  \ref{DandWo}{Dowker,J.S. and Wolski, A. \prA{46}{1992}{6417}.}
  \ref{Zeitlin1}{Zeitlin,V. {\it Physica D} {\bf 49} (1991).  }
  \ref{Zeitlin0}{Zeitlin,V. {\it Nonlinear World} Ed by
   V.Baryakhtar {\it et al},  Vol.I p.717,  (World Scientific, Singapore, 1989).}
  \ref{Zeitlin2}{Zeitlin,V. \prl{93}{2004}{264501}. }
  \ref{Zeitlin3}{Zeitlin,V. \pla{339}{2005}{316}. }
  \ref{Groenewold}{Groenewold, H.J. {\it Physica} {\bf 12} (1946) 405.}
  \ref{Cohen}{Cohen, L. \jmp{7}{1966}{781}.}
  \ref{AandW}{Argawal G.S. and Wolf, E. \prD{2}{1970}{2161,2187,2206}.}
  \ref{Jantzen}{Jantzen,R.T. \jmp{19}{1978}{1163}.}
  \ref{Moses2}{Moses,H.E. \aop{42}{1967}{343}.}
  \ref{Carmeli}{Carmeli,M. \jmp{9}{1968}{1987}.}
  \ref{SHS}{Siemans,M., Hancock,J. and Siminovitch,D. {\it Solid State
  Nuclear Magnetic Resonance} {\bf 31}(2007)35.}
 \ref{Dowk}{Dowker,J.S. \prD{28}{1983}{3013}.}
 \ref{Heine}{Heine, E. {\it Handbuch der Kugelfunctionen}
  (G.Reimer, Berlin. 1878, 1881).}
  \ref{Pockels}{Pockels, F. {\it \"Uber die Differentialgleichung $\De
  u+k^2u=0$} (Teubner, Leipzig. 1891).}
  \ref{Hamermesh}{Hamermesh, M., {\it Group Theory} (Addison--Wesley,
  Reading. 1962).}
  \ref{Racah}{Racah, G. {\it Group Theory and Spectroscopy}
  (Princeton Lecture Notes, 1951). }
  \ref{Gourdin}{Gourdin, M. {\it Basics of Lie Groups} (Editions
  Fronti\'eres, Gif sur Yvette. 1982.)}
  \ref{Clifford}{Clifford, W.K. \plms{2}{1866}{116}.}
  \ref{Story2}{Story, W.E. \plms{23}{1892}{265}.}
  \ref{Story}{Story, W.E. \ma{41}{1893}{469}.}
  \ref{Poole}{Poole, E.G.C. \plms{33}{1932}{435}.}
  \ref{Dickson}{Dickson, L.E. {\it Algebraic Invariants} (Wiley, N.Y.
  1915).}
  \ref{Dickson2}{Dickson, L.E. {\it Modern Algebraic Theories}
  (Sanborn and Co., Boston. 1926).}
  \ref{Hilbert2}{Hilbert, D. {\it Theory of algebraic invariants} (C.U.P.,
  Cambridge. 1993).}
  \ref{Olver}{Olver, P.J. {\it Classical Invariant Theory} (C.U.P., Cambridge.
  1999.)}
  \ref{AST}{A\v{s}erova, R.M., Smirnov, J.F. and Tolsto\v{i}, V.N. {\it
  Teoret. Mat. Fyz.} {\bf 8} (1971) 255.}
  \ref{AandS}{A\v{s}erova, R.M., Smirnov, J.F. \np{4}{1968}{399}.}
  \ref{Shapiro}{Shapiro, J. \jmp{6}{1965}{1680}.}
  \ref{Shapiro2}{Shapiro, J.Y. \jmp{14}{1973}{1262}.}
  \ref{NandS}{Noz, M.E. and Shapiro, J.Y. \np{51}{1973}{309}.}
  \ref{Cayley2}{Cayley, A. {\it Phil. Trans. Roy. Soc. Lond.}
  {\bf 144} (1854) 244.}
  \ref{Cayley3}{Cayley, A. {\it Phil. Trans. Roy. Soc. Lond.}
  {\bf 146} (1856) 101.}
  \ref{Wigner}{Wigner, E.P. {\it Gruppentheorie} (Vieweg, Braunschweig. 1931).}
  \ref{Sharp}{Sharp, R.T. \ajop{28}{1960}{116}.}
  \ref{Laporte}{Laporte, O. {\it Z. f. Naturf.} {\bf 3a} (1948) 447.}
  \ref{Lowdin}{L\"owdin, P-O. \rmp{36}{1964}{966}.}
  \ref{Ansari}{Ansari, S.M.R. {\it Fort. d. Phys.} {\bf 15} (1967) 707.}
  \ref{SSJR}{Samal, P.K., Saha, R., Jain, P. and Ralston, J.P. {\it
  Testing Isotropy of Cosmic Microwave Background Radiation},
  astro-ph/0708.2816.}
  \ref{Lachieze}{Lachi\'eze-Rey, M. {\it Harmonic projection and
  multipole Vectors}. astro- \break ph/0409081.}
  \ref{CHS}{Copi, C.J., Huterer, D. and Starkman, G.D.
  \prD{70}{2003}{043515}.}
  \ref{Jaric}{Jari\'c, J.P. {\it Int. J. Eng. Sci.} {\bf 41} (2003) 2123.}
  \ref{RandD}{Roche, J.A. and Dowker, J.S. \jpa{1}{1968}{527}.}
  \ref{KandW}{Katz, G. and Weeks, J.R. \prD{70}{2004}{063527}.}
  \ref{Waerden}{van der Waerden, B.L. {\it Die Gruppen-theoretische
  Methode in der Quantenmechanik} (Springer, Berlin. 1932).}
  \ref{EMOT}{Erdelyi, A., Magnus, W., Oberhettinger, F. and Tricomi, F.G. {
  \it Higher Transcendental Functions} Vol.1 (McGraw-Hill, N.Y. 1953).}
  \ref{Dowzilch}{Dowker, J.S. {\it Proc. Phys. Soc.} {\bf 91} (1967) 28.}
  \ref{DandD}{Dowker, J.S. and Dowker, Y.P. {\it Proc. Phys. Soc.}
  {\bf 87} (1966) 65.}
  \ref{DandD2}{Dowker, J.S. and Dowker, Y.P. \prs{}{}{}.}
  \ref{Dowk3}{Dowker,J.S. \cqg{7}{1990}{1241}.}
  \ref{Dowk5}{Dowker,J.S. \cqg{7}{1990}{2353}.}
  \ref{CoandH}{Courant, R. and Hilbert, D. {\it Methoden der
  Mathematischen Physik} vol.1 \break (Springer, Berlin. 1931).}
  \ref{Applequist}{Applequist, J. \jpa{22}{1989}{4303}.}
  \ref{Torruella}{Torruella, \jmp{16}{1975}{1637}.}
  \ref{Weinberg}{Weinberg, S.W. \pr{133}{1964}{B1318}.}
  \ref{Meyerw}{Meyer, W.F. {\it Apolarit\"at und rationale Curven}
  (Fues, T\"ubingen. 1883.) }
  \ref{Ostrowski}{Ostrowski, A. {\it Jahrsb. Deutsch. Math. Verein.} {\bf
  33} (1923) 245.}
  \ref{Kramers}{Kramers, H.A. {\it Grundlagen der Quantenmechanik}, (Akad.
  Verlag., Leipzig, 1938).}
  \ref{ZandZ}{Zou, W.-N. and Zheng, Q.-S. \prs{459}{2003}{527}.}
  \ref{Weeks1}{Weeks, J.R. {\it Maxwell's multipole vectors
  and the CMB}.  astro-ph/0412231.}
  \ref{Corson}{Corson, E.M. {\it Tensors, Spinors and Relativistic Wave
  Equations} (Blackie, London. 1950).}
  \ref{Rosanes}{Rosanes, J. \jram{76}{1873}{312}.}
  \ref{Salmon}{Salmon, G. {\it Lessons Introductory to the Modern Higher
  Algebra} 3rd. edn. \break (Hodges,  Dublin. 1876.)}
  \ref{Milnew}{Milne, W.P. {\it Homogeneous Coordinates} (Arnold. London. 1910).}
  \ref{Niven}{Niven, W.D. {\it Phil. Trans. Roy. Soc.} {\bf 170} (1879) 393.}
  \ref{Scott}{Scott, C.A. {\it An Introductory Account of
  Certain Modern Ideas and Methods in Plane Analytical Geometry,}
  (MacMillan, N.Y. 1896).}
  \ref{Bargmann}{Bargmann, V. \rmp{34}{1962}{300}.}
  \ref{Maxwell}{Maxwell, J.C. {\it A Treatise on Electricity and
  Magnetism} 2nd. edn. (Clarendon Press, Oxford. 1882).}
  \ref{BandL}{Biedenharn, L.C. and Louck, J.D.
  {\it Angular Momentum in Quantum Physics} (Addison-Wesley, Reading. 1981).}
  \ref{Weylqm}{Weyl, H. {\it The Theory of Groups and Quantum Mechanics}
  (Methuen, London. 1931).}
  \ref{Robson}{Robson, A. {\it An Introduction to Analytical Geometry} Vol I
  (C.U.P., Cambridge. 1940.)}
  \ref{Sommerville}{Sommerville, D.M.Y. {\it Analytical Conics} 3rd. edn.
   (Bell, London. 1933).}
  \ref{Coolidge}{Coolidge, J.L. {\it A Treatise on Algebraic Plane Curves}
  (Clarendon Press, Oxford. 1931).}
  \ref{SandK}{Semple, G. and Kneebone. G.T. {\it Algebraic Projective
  Geometry} (Clarendon Press, Oxford. 1952).}
  \ref{AandC}{Abdesselam A., and Chipalkatti, J. {\it The Higher
  Transvectants are redundant}, arXiv:0801.1533 [math.AG] 2008.}
  \ref{Elliott}{Elliott, E.B. {\it The Algebra of Quantics} 2nd edn.
  (Clarendon Press, Oxford. 1913).}
  \ref{Elliott2}{Elliott, E.B. \qjpam{48}{1917}{372}.}
  \ref{Howe}{Howe, R. \tams{313}{1989}{539}.}
  \ref{Clebsch}{Clebsch, A. \jram{60}{1862}{343}.}
  \ref{Prasad}{Prasad, G. \ma{72}{1912}{136}.}
  \ref{Dougall}{Dougall, J. \pems{32}{1913}{30}.}
  \ref{Penrose}{Penrose, R. \aop{10}{1960}{171}.}
  \ref{Penrose2}{Penrose, R. \prs{273}{1965}{171}.}
  \ref{Burnside}{Burnside, W.S. \qjm{10}{1870}{211}. }
  \ref{Lindemann}{Lindemann, F. \ma{23} {1884}{111}.}
  \ref{Backus}{Backus, G. {\it Rev. Geophys. Space Phys.} {\bf 8} (1970) 633.}
  \ref{Baerheim}{Baerheim, R. {\it Q.J. Mech. appl. Math.} {\bf 51} (1998) 73.}
  \ref{Lense}{Lense, J. {\it Kugelfunktionen} (Akad.Verlag, Leipzig. 1950).}
  \ref{Littlewood}{Littlewood, D.E. \plms{50}{1948}{349}.}
  \ref{Fierz}{Fierz, M. {\it Helv. Phys. Acta} {\bf 12} (1938) 3.}
  \ref{Williams}{Williams, D.N. {\it Lectures in Theoretical Physics} Vol. VII,
  (Univ.Colorado Press, Boulder. 1965).}
  \ref{Dennis}{Dennis, M. \jpa{37}{2004}{9487}.}
  \ref{Pirani}{Pirani, F. {\it Brandeis Lecture Notes on
  General Relativity,} edited by S. Deser and K. Ford. (Brandeis, Mass. 1964).}
  \ref{Sturm}{Sturm, R. \jram{86}{1878}{116}.}
  \ref{Schlesinger}{Schlesinger, O. \ma{22}{1883}{521}.}
  \ref{Askwith}{Askwith, E.H. {\it Analytical Geometry of the Conic
  Sections} (A.\&C. Black, London. 1908).}
  \ref{Todd}{Todd, J.A. {\it Projective and Analytical Geometry}.
  (Pitman, London. 1946).}
  \ref{Glenn}{Glenn. O.E. {\it Theory of Invariants} (Ginn \& Co, N.Y. 1915).}
  \ref{DowkandG}{Dowker, J.S. and Goldstone, M. \prs{303}{1968}{381}.}
  \ref{Turnbull}{Turnbull, H.A. {\it The Theory of Determinants,
  Matrices and Invariants} 3rd. edn. (Dover, N.Y. 1960).}
  \ref{MacMillan}{MacMillan, W.D. {\it The Theory of the Potential}
  (McGraw-Hill, N.Y. 1930).}
   \ref{Hobson}{Hobson, E.W. {\it The Theory of Spherical
   and Ellipsoidal Harmonics} (C.U.P., Cambridge. 1931).}
  \ref{Hobson1}{Hobson, E.W. \plms {24}{1892}{55}.}
  \ref{GandY}{Grace, J.H. and Young, A. {\it The Algebra of Invariants}
  (C.U.P., Cambridge, 1903).}
  \ref{FandR}{Fano, U. and Racah, G. {\it Irreducible Tensorial Sets}
  (Academic Press, N.Y. 1959).}
  \ref{TandT}{Thomson, W. and Tait, P.G. {\it Treatise on Natural Philosophy}
  (Clarendon Press, Oxford. 1867).}
  \ref{Brinkman}{Brinkman, H.C. {\it Applications of spinor invariants in
atomic physics}, North Holland, Amsterdam 1956.}
  \ref{Kramers1}{Kramers, H.A. {\it Proc. Roy. Soc. Amst.} {\bf 33} (1930) 953.}
  \ref{DandP2}{Dowker,J.S. and Pettengill,D.F. \jpa{7}{1974}{1527}}
  \ref{Dowk1}{Dowker,J.S. \jpa{}{}{45}.}
  \ref{Dowk2}{Dowker,J.S. \aop{71}{1972}{577}}
  \ref{DandA}{Dowker,J.S. and Apps, J.S. \cqg{15}{1998}{1121}.}
  \ref{Weil}{Weil,A., {\it Elliptic functions according to Eisenstein
  and Kronecker}, Springer, Berlin, 1976.}
  \ref{Ling}{Ling,C-H. {\it SIAM J.Math.Anal.} {\bf5} (1974) 551.}
  \ref{Ling2}{Ling,C-H. {\it J.Math.Anal.Appl.}(1988).}
 \ref{BMO}{Brevik,I., Milton,K.A. and Odintsov, S.D. \aop{302}{2002}{120}.}
 \ref{KandL}{Kutasov,D. and Larsen,F. {\it JHEP} 0101 (2001) 1.}
 \ref{KPS}{Klemm,D., Petkou,A.C. and Siopsis {\it Entropy
 bounds, monoticity properties and scaling in CFT's}. hep-th/0101076.}
 \ref{DandC}{Dowker,J.S. and Critchley,R. \prD{15}{1976}{1484}.}
 \ref{AandD}{Al'taie, M.B. and Dowker, J.S. \prD{18}{1978}{3557}.}
 \ref{Dow1}{Dowker,J.S. \prD{37}{1988}{558}.}
 \ref{Dow30}{Dowker,J.S. \prD{28}{1983}{3013}.}
 \ref{DandK}{Dowker,J.S. and Kennedy,G. \jpa{}{1978}{}.}
 \ref{Dow2}{Dowker,J.S. \cqg{1}{1984}{359}.}
 \ref{DandKi}{Dowker,J.S. and Kirsten, K. {\it Comm. in Anal. and Geom.
 }{\bf7} (1999) 641.}
 \ref{DandKe}{Dowker,J.S. and Kennedy,G.\jpa{11}{1978}{895}.}
 \ref{Gibbons}{Gibbons,G.W. \pl{60A}{1977}{385}.}
 \ref{Cardy}{Cardy,J.L. \np{366}{1991}{403}.}
 \ref{ChandD}{Chang,P. and Dowker,J.S. \np{395}{1993}{407}.}
 \ref{DandC2}{Dowker,J.S. and Critchley,R. \prD{13}{1976}{224}.}
 \ref{Camporesi}{Camporesi,R. \prp{196}{1990}{1}.}
 \ref{BandM}{Brown,L.S. and Maclay,G.J. \pr{184}{1969}{1272}.}
 \ref{CandD}{Candelas,P. and Dowker,J.S. \prD{19}{1979}{2902}.}
 \ref{Unwin1}{Unwin,S.D. Thesis. University of Manchester. 1979.}
 \ref{Unwin2}{Unwin,S.D. \jpa{13}{1980}{313}.}
 \ref{DandB}{Dowker,J.S.and Banach,R. \jpa{11}{1978}{2255}.}
 \ref{Obhukov}{Obhukov,Yu.N. \pl{109B}{1982}{195}.}
 \ref{Kennedy}{Kennedy,G. \prD{23}{1981}{2884}.}
 \ref{CandT}{Copeland,E. and Toms,D.J. \np {255}{1985}{201}.}
 \ref{ELV}{Elizalde,E., Lygren, M. and Vassilevich,
 D.V. \jmp {37}{1996}{3105}.}
 \ref{Malurkar}{Malurkar,S.L. {\it J.Ind.Math.Soc} {\bf16} (1925/26) 130.}
 \ref{Glaisher}{Glaisher,J.W.L. {\it Messenger of Math.} {\bf18}
(1889) 1.} \ref{Anderson}{Anderson,A. \prD{37}{1988}{536}.}
 \ref{CandA}{Cappelli,A. and D'Appollonio,\pl{487B}{2000}{87}.}
 \ref{Wot}{Wotzasek,C. \jpa{23}{1990}{1627}.}
 \ref{RandT}{Ravndal,F. and Tollesen,D. \prD{40}{1989}{4191}.}
 \ref{SandT}{Santos,F.C. and Tort,A.C. \pl{482B}{2000}{323}.}
 \ref{FandO}{Fukushima,K. and Ohta,K. {\it Physica} {\bf A299} (2001) 455.}
 \ref{GandP}{Gibbons,G.W. and Perry,M. \prs{358}{1978}{467}.}
 \ref{Dow4}{Dowker,J.S..}
  \ref{Rad}{Rademacher,H. {\it Topics in analytic number theory,}
Springer-Verlag,  Berlin,1973.}
  \ref{Halphen}{Halphen,G.-H. {\it Trait\'e des Fonctions Elliptiques},
  Vol 1, Gauthier-Villars, Paris, 1886.}
  \ref{CandW}{Cahn,R.S. and Wolf,J.A. {\it Comm.Mat.Helv.} {\bf 51}
  (1976) 1.}
  \ref{Berndt}{Berndt,B.C. \rmjm{7}{1977}{147}.}
  \ref{Hurwitz}{Hurwitz,A. \ma{18}{1881}{528}.}
  \ref{Hurwitz2}{Hurwitz,A. {\it Mathematische Werke} Vol.I. Basel,
  Birkhauser, 1932.}
  \ref{Berndt2}{Berndt,B.C. \jram{303/304}{1978}{332}.}
  \ref{RandA}{Rao,M.B. and Ayyar,M.V. \jims{15}{1923/24}{150}.}
  \ref{Hardy}{Hardy,G.H. \jlms{3}{1928}{238}.}
  \ref{TandM}{Tannery,J. and Molk,J. {\it Fonctions Elliptiques},
   Gauthier-Villars, Paris, 1893--1902.}
  \ref{schwarz}{Schwarz,H.-A. {\it Formeln und
  Lehrs\"atzen zum Gebrauche..},Springer 1893.(The first edition was 1885.)
  The French translation by Henri Pad\'e is {\it Formules et Propositions
  pour L'Emploi...},Gauthier-Villars, Paris, 1894}
  \ref{Hancock}{Hancock,H. {\it Theory of elliptic functions}, Vol I.
   Wiley, New York 1910.}
  \ref{watson}{Watson,G.N. \jlms{3}{1928}{216}.}
  \ref{MandO}{Magnus,W. and Oberhettinger,F. {\it Formeln und S\"atze},
  Springer-Verlag, Berlin 1948.}
  \ref{Klein}{Klein,F. {\it Lectures on the Icosohedron}
  (Methuen, London. 1913).}
  \ref{AandL}{Appell,P. and Lacour,E. {\it Fonctions Elliptiques},
  Gauthier-Villars,
  Paris. 1897.}
  \ref{HandC}{Hurwitz,A. and Courant,C. {\it Allgemeine Funktionentheorie},
  Springer,
  Berlin. 1922.}
  \ref{WandW}{Whittaker,E.T. and Watson,G.N. {\it Modern analysis},
  Cambridge. 1927.}
  \ref{SandC}{Selberg,A. and Chowla,S. \jram{227}{1967}{86}. }
  \ref{zucker}{Zucker,I.J. {\it Math.Proc.Camb.Phil.Soc} {\bf 82 }(1977)
  111.}
  \ref{glasser}{Glasser,M.L. {\it Maths.of Comp.} {\bf 25} (1971) 533.}
  \ref{GandW}{Glasser, M.L. and Wood,V.E. {\it Maths of Comp.} {\bf 25}
  (1971)
  535.}
  \ref{greenhill}{Greenhill,A,G. {\it The Applications of Elliptic
  Functions}, MacMillan. London, 1892.}
  \ref{Weierstrass}{Weierstrass,K. {\it J.f.Mathematik (Crelle)}
{\bf 52} (1856) 346.}
  \ref{Weierstrass2}{Weierstrass,K. {\it Mathematische Werke} Vol.I,p.1,
  Mayer u. M\"uller, Berlin, 1894.}
  \ref{Fricke}{Fricke,R. {\it Die Elliptische Funktionen und Ihre Anwendungen},
    Teubner, Leipzig. 1915, 1922.}
  \ref{Konig}{K\"onigsberger,L. {\it Vorlesungen \"uber die Theorie der
 Elliptischen Funktionen},  \break Teubner, Leipzig, 1874.}
  \ref{Milne}{Milne,S.C. {\it The Ramanujan Journal} {\bf 6} (2002) 7-149.}
  \ref{Schlomilch}{Schl\"omilch,O. {\it Ber. Verh. K. Sachs. Gesell. Wiss.
  Leipzig}  {\bf 29} (1877) 101-105; {\it Compendium der h\"oheren
  Analysis}, Bd.II, 3rd Edn, Vieweg, Brunswick, 1878.}
  \ref{BandB}{Briot,C. and Bouquet,C. {\it Th\`eorie des Fonctions
  Elliptiques}, Gauthier-Villars, Paris, 1875.}
  \ref{Dumont}{Dumont,D. \aim {41}{1981}{1}.}
  \ref{Andre}{Andr\'e,D. {\it Ann.\'Ecole Normale Superior} {\bf 6} (1877)
  265;
  {\it J.Math.Pures et Appl.} {\bf 5} (1878) 31.}
  \ref{Raman}{Ramanujan,S. {\it Trans.Camb.Phil.Soc.} {\bf 22} (1916) 159;
 {\it Collected Papers}, Cambridge, 1927}
  \ref{Weber}{Weber,H.M. {\it Lehrbuch der Algebra} Bd.III, Vieweg,
  Brunswick 190  3.}
  \ref{Weber2}{Weber,H.M. {\it Elliptische Funktionen und algebraische
  Zahlen},
  Vieweg, Brunswick 1891.}
  \ref{ZandR}{Zucker,I.J. and Robertson,M.M.
  {\it Math.Proc.Camb.Phil.Soc} {\bf 95 }(1984) 5.}
  \ref{JandZ1}{Joyce,G.S. and Zucker,I.J.
  {\it Math.Proc.Camb.Phil.Soc} {\bf 109 }(1991) 257.}
  \ref{JandZ2}{Zucker,I.J. and Joyce.G.S.
  {\it Math.Proc.Camb.Phil.Soc} {\bf 131 }(2001) 309.}
  \ref{zucker2}{Zucker,I.J. {\it SIAM J.Math.Anal.} {\bf 10} (1979) 192,}
  \ref{BandZ}{Borwein,J.M. and Zucker,I.J. {\it IMA J.Math.Anal.} {\bf 12}
  (1992) 519.}
  \ref{Cox}{Cox,D.A. {\it Primes of the form $x^2+n\,y^2$}, Wiley,
  New York, 1989.}
  \ref{BandCh}{Berndt,B.C. and Chan,H.H. {\it Mathematika} {\bf42} (1995)
  278.}
  \ref{EandT}{Elizalde,R. and Tort.hep-th/}
  \ref{KandS}{Kiyek,K. and Schmidt,H. {\it Arch.Math.} {\bf 18} (1967) 438.}
  \ref{Oshima}{Oshima,K. \prD{46}{1992}{4765}.}
  \ref{greenhill2}{Greenhill,A.G. \plms{19} {1888} {301}.}
  \ref{Russell}{Russell,R. \plms{19} {1888} {91}.}
  \ref{BandB}{Borwein,J.M. and Borwein,P.B. {\it Pi and the AGM}, Wiley,
  New York, 1998.}
  \ref{Resnikoff}{Resnikoff,H.L. \tams{124}{1966}{334}.}
  \ref{vandp}{Van der Pol, B. {\it Indag.Math.} {\bf18} (1951) 261,272.}
  \ref{Rankin}{Rankin,R.A. {\it Modular forms} C.U.P. Cambridge}
  \ref{Rankin2}{Rankin,R.A. {\it Proc. Roy.Soc. Edin.} {\bf76 A} (1976) 107.}
  \ref{Skoruppa}{Skoruppa,N-P. {\it J.of Number Th.} {\bf43} (1993) 68 .}
  \ref{Down}{Dowker.J.S. \np {104}{2002}{153}.}
  \ref{Eichler}{Eichler,M. \mz {67}{1957}{267}.}
  \ref{Zagier}{Zagier,D. \invm{104}{1991}{449}.}
  \ref{Lang}{Lang,S. {\it Modular Forms}, Springer, Berlin, 1976.}
  \ref{Kosh}{Koshliakov,N.S. {\it Mess.of Math.} {\bf 58} (1928) 1.}
  \ref{BandH}{Bodendiek, R. and Halbritter,U. \amsh{38}{1972}{147}.}
  \ref{Smart}{Smart,L.R., \pgma{14}{1973}{1}.}
  \ref{Grosswald}{Grosswald,E. {\it Acta. Arith.} {\bf 21} (1972) 25.}
  \ref{Kata}{Katayama,K. {\it Acta Arith.} {\bf 22} (1973) 149.}
  \ref{Ogg}{Ogg,A. {\it Modular forms and Dirichlet series} (Benjamin,
  New York,
   1969).}
  \ref{Bol}{Bol,G. \amsh{16}{1949}{1}.}
  \ref{Epstein}{Epstein,P. \ma{56}{1903}{615}.}
  \ref{Petersson}{Petersson.}
  \ref{Serre}{Serre,J-P. {\it A Course in Arithmetic}, Springer,
  New York, 1973.}
  \ref{Schoenberg}{Schoenberg,B., {\it Elliptic Modular Functions},
  Springer, Berlin, 1974.}
  \ref{Apostol}{Apostol,T.M. \dmj {17}{1950}{147}.}
  \ref{Ogg2}{Ogg,A. {\it Lecture Notes in Math.} {\bf 320} (1973) 1.}
  \ref{Knopp}{Knopp,M.I. \dmj {45}{1978}{47}.}
  \ref{Knopp2}{Knopp,M.I. \invm {}{1994}{361}.}
  \ref{LandZ}{Lewis,J. and Zagier,D. \aom{153}{2001}{191}.}
  \ref{DandK1}{Dowker,J.S. and Kirsten,K. {\it Elliptic functions and
  temperature inversion symmetry on spheres} hep-th/.}
  \ref{HandK}{Husseini and Knopp.}
  \ref{Kober}{Kober,H. \mz{39}{1934-5}{609}.}
  \ref{HandL}{Hardy,G.H. and Littlewood, \am{41}{1917}{119}.}
  \ref{Watson}{Watson,G.N. \qjm{2}{1931}{300}.}
  \ref{SandC2}{Chowla,S. and Selberg,A. {\it Proc.Nat.Acad.} {\bf 35}
  (1949) 371.}
  \ref{Landau}{Landau, E. {\it Lehre von der Verteilung der Primzahlen},
  (Teubner, Leipzig, 1909).}
  \ref{Berndt4}{Berndt,B.C. \tams {146}{1969}{323}.}
  \ref{Berndt3}{Berndt,B.C. \tams {}{}{}.}
  \ref{Bochner}{Bochner,S. \aom{53}{1951}{332}.}
  \ref{Weil2}{Weil,A.\ma{168}{1967}{}.}
  \ref{CandN}{Chandrasekharan,K. and Narasimhan,R. \aom{74}{1961}{1}.}
  \ref{Rankin3}{Rankin,R.A. {} {} ().}
  \ref{Berndt6}{Berndt,B.C. {\it Trans.Edin.Math.Soc}.}
  \ref{Elizalde}{Elizalde,E. {\it Ten Physical Applications of Spectral
  Zeta Function Theory}, \break (Springer, Berlin, 1995).}
  \ref{Allen}{Allen,B., Folacci,A. and Gibbons,G.W. \pl{189}{1987}{304}.}
  \ref{Krazer}{Krazer}
  \ref{Elizalde3}{Elizalde,E. {\it J.Comp.and Appl. Math.} {\bf 118}
  (2000) 125.}
  \ref{Elizalde2}{Elizalde,E., Odintsov.S.D, Romeo, A. and Bytsenko,
  A.A and
  Zerbini,S.
  {\it Zeta function regularisation}, (World Scientific, Singapore,
  1994).}
  \ref{Eisenstein}{Eisenstein}
  \ref{Hecke}{Hecke,E. \ma{112}{1936}{664}.}
  \ref{Hecke2}{Hecke,E. \ma{112}{1918}{398}.}
  \ref{Terras}{Terras,A. {\it Harmonic analysis on Symmetric Spaces} (Springer,
  New York, 1985).}
  \ref{BandG}{Bateman,P.T. and Grosswald,E. {\it Acta Arith.} {\bf 9}
  (1964) 365.}
  \ref{Deuring}{Deuring,M. \aom{38}{1937}{585}.}
  \ref{Guinand}{Guinand.}
  \ref{Guinand2}{Guinand.}
  \ref{Minak}{Minakshisundaram.}
  \ref{Mordell}{Mordell,J. \prs{}{}{}.}
  \ref{GandZ}{Glasser,M.L. and Zucker, {}.}
  \ref{Landau2}{Landau,E. \jram{}{1903}{64}.}
  \ref{Kirsten1}{Kirsten,K. \jmp{35}{1994}{459}.}
  \ref{Sommer}{Sommer,J. {\it Vorlesungen \"uber Zahlentheorie}
  (1907,Teubner,Leipzig).
  French edition 1913 .}
  \ref{Reid}{Reid,L.W. {\it Theory of Algebraic Numbers},
  (1910,MacMillan,New York).}
  \ref{Milnor}{Milnor, J. {\it Is the Universe simply--connected?},
  IAS, Princeton, 1978.}
  \ref{Milnor2}{Milnor, J. \ajm{79}{1957}{623}.}
  \ref{Opechowski}{Opechowski,W. {\it Physica} {\bf 7} (1940) 552.}
  \ref{Bethe}{Bethe, H.A. \zfp{3}{1929}{133}.}
  \ref{LandL}{Landau, L.D. and Lishitz, E.M. {\it Quantum
  Mechanics} (Pergamon Press, London, 1958).}
  \ref{GPR}{Gibbons, G.W., Pope, C. and R\"omer, H., \np{157}{1979}{377}.}
  \ref{Jadhav}{Jadhav,S.P. PhD Thesis, University of Manchester 1990.}
  \ref{DandJ}{Dowker,J.S. and Jadhav, S. \prD{39}{1989}{1196}.}
  \ref{CandM}{Coxeter, H.S.M. and Moser, W.O.J. {\it Generators and
  relations of finite groups} (Springer. Berlin. 1957).}
  \ref{Coxeter2}{Coxeter, H.S.M. {\it Regular Complex Polytopes},
   (Cambridge University Press, \break Cambridge, 1975).}
  \ref{Coxeter}{Coxeter, H.S.M. {\it Regular Polytopes}.}
  \ref{Stiefel}{Stiefel, E., J.Research NBS {\bf 48} (1952) 424.}
  \ref{BandS}{Brink, D.M. and Satchler, G.R. {\it Angular momentum theory}.
  (Clarendon Press, Oxford. 1962.).}
  \ref{Rose}{Rose}
  \ref{Schwinger}{Schwinger, J. {\it On Angular Momentum}
  in {\it Quantum Theory of Angular Momentum} edited by
  Biedenharn,L.C. and van Dam, H. (Academic Press, N.Y. 1965).}
  \ref{Bromwich}{Bromwich, T.J.I'A. {\it Infinite Series},
  (Macmillan, 1947).}
  \ref{Ray}{Ray,D.B. \aim{4}{1970}{109}.}
  \ref{Ikeda}{Ikeda,A. {\it Kodai Math.J.} {\bf 18} (1995) 57.}
  \ref{Kennedy}{Kennedy,G. \prD{23}{1981}{2884}.}
  \ref{Ellis}{Ellis,G.F.R. {\it General Relativity} {\bf2} (1971) 7.}
  \ref{Dow8}{Dowker,J.S. \cqg{20}{2003}{L105}.}
  \ref{IandY}{Ikeda, A and Yamamoto, Y. \ojm {16}{1979}{447}.}
  \ref{BandI}{Bander,M. and Itzykson,C. \rmp{18}{1966}{2}.}
  \ref{Schulman}{Schulman, L.S. \pr{176}{1968}{1558}.}
  \ref{Bar1}{B\"ar,C. {\it Arch.d.Math.}{\bf 59} (1992) 65.}
  \ref{Bar2}{B\"ar,C. {\it Geom. and Func. Anal.} {\bf 6} (1996) 899.}
  \ref{Vilenkin}{Vilenkin, N.J. {\it Special functions},
  (Am.Math.Soc., Providence, 1968).}
  \ref{Talman}{Talman, J.D. {\it Special functions} (Benjamin,N.Y.,1968).}
  \ref{Miller}{Miller, W. {\it Symmetry groups and their applications}
  (Wiley, N.Y., 1972).}
  \ref{Dow3}{Dowker,J.S. \cmp{162}{1994}{633}.}
  \ref{Cheeger}{Cheeger, J. \jdg {18}{1983}{575}.}
  \ref{Cheeger2}{Cheeger, J. \aom {109}{1979}{259}.}
  \ref{Dow6}{Dowker,J.S. \jmp{30}{1989}{770}.}
  \ref{Dow20}{Dowker,J.S. \jmp{35}{1994}{6076}.}
  \ref{Dowjmp}{Dowker,J.S. \jmp{35}{1994}{4989}.}
  \ref{Dow21}{Dowker,J.S. {\it Heat kernels and polytopes} in {\it
   Heat Kernel Techniques and Quantum Gravity}, ed. by S.A.Fulling,
   Discourses in Mathematics and its Applications, No.4, Dept.
   Maths., Texas A\&M University, College Station, Texas, 1995.}
  \ref{Dow9}{Dowker,J.S. \jmp{42}{2001}{1501}.}
  \ref{Dow7}{Dowker,J.S. \jpa{25}{1992}{2641}.}
  \ref{Warner}{Warner.N.P. \prs{383}{1982}{379}.}
  \ref{Wolf}{Wolf, J.A. {\it Spaces of constant curvature},
  (McGraw--Hill,N.Y., 1967).}
  \ref{Meyer}{Meyer,B. \cjm{6}{1954}{135}.}
  \ref{BandB}{B\'erard,P. and Besson,G. {\it Ann. Inst. Four.} {\bf 30}
  (1980) 237.}
  \ref{PandM}{Polya,G. and Meyer,B. \cras{228}{1948}{28}.}
  \ref{Springer}{Springer, T.A. Lecture Notes in Math. vol 585 (Springer,
  Berlin,1977).}
  \ref{SeandT}{Threlfall, H. and Seifert, W. \ma{104}{1930}{1}.}
  \ref{Hopf}{Hopf,H. \ma{95}{1925}{313}. }
  \ref{Dow}{Dowker,J.S. \jpa{5}{1972}{936}.}
  \ref{LLL}{Lehoucq,R., Lachi\'eze-Rey,M. and Luminet, J.--P. {\it
  Astron.Astrophys.} {\bf 313} (1996) 339.}
  \ref{LaandL}{Lachi\'eze-Rey,M. and Luminet, J.--P.
  \prp{254}{1995}{135}.}
  \ref{Schwarzschild}{Schwarzschild, K., {\it Vierteljahrschrift der
  Ast.Ges.} {\bf 35} (1900) 337.}
  \ref{Starkman}{Starkman,G.D. \cqg{15}{1998}{2529}.}
  \ref{LWUGL}{Lehoucq,R., Weeks,J.R., Uzan,J.P., Gausman, E. and
  Luminet, J.--P. \cqg{19}{2002}{4683}.}
  \ref{Dow10}{Dowker,J.S. \prD{28}{1983}{3013}.}
  \ref{BandD}{Banach, R. and Dowker, J.S. \jpa{12}{1979}{2527}.}
  \ref{Jadhav2}{Jadhav,S. \prD{43}{1991}{2656}.}
  \ref{Gilkey}{Gilkey,P.B. {\it Invariance theory,the heat equation and
  the Atiyah--Singer Index theorem} (CRC Press, Boca Raton, 1994).}
  \ref{BandY}{Berndt,B.C. and Yeap,B.P. {\it Adv. Appl. Math.}
  {\bf29} (2002) 358.}
  \ref{HandR}{Hanson,A.J. and R\"omer,H. \pl{80B}{1978}{58}.}
  \ref{Hill}{Hill,M.J.M. {\it Trans.Camb.Phil.Soc.} {\bf 13} (1883) 36.}
  \ref{Cayley}{Cayley,A. {\it Quart.Math.J.} {\bf 7} (1866) 304.}
  \ref{Seade}{Seade,J.A. {\it Anal.Inst.Mat.Univ.Nac.Aut\'on
  M\'exico} {\bf 21} (1981) 129.}
  \ref{CM}{Cisneros--Molina,J.L. {\it Geom.Dedicata} {\bf84} (2001)
  \ref{Goette1}{Goette,S. \jram {526} {2000} 181.}
  207.}
  \ref{NandO}{Nash,C. and O'Connor,D--J, \jmp {36}{1995}{1462}.}
  \ref{Dows}{Dowker,J.S. \aop{71}{1972}{577}; Dowker,J.S. and Pettengill,D.F.
  \jpa{7}{1974}{1527}; J.S.Dowker in {\it Quantum Gravity}, edited by
  S. C. Christensen (Hilger,Bristol,1984)}
  \ref{Jadhav2}{Jadhav,S.P. \prD{43}{1991}{2656}.}
  \ref{Dow11}{Dowker,J.S. \cqg{21}{2004}4247.}
  \ref{Dow12}{Dowker,J.S. \cqg{21}{2004}4977.}
  \ref{Dow13}{Dowker,J.S. \jpa{38}{2005}1049.}
  \ref{Zagier}{Zagier,D. \ma{202}{1973}{149}}
  \ref{RandG}{Rademacher, H. and Grosswald,E. {\it Dedekind Sums},
  (Carus, MAA, 1972).}
  \ref{Berndt7}{Berndt,B, \aim{23}{1977}{285}.}
  \ref{HKMM}{Harvey,J.A., Kutasov,D., Martinec,E.J. and Moore,G.
  {\it Localised Tachyons and RG Flows}, hep-th/0111154.}
  \ref{Beck}{Beck,M., {\it Dedekind Cotangent Sums}, {\it Acta Arithmetica}
  {\bf 109} (2003) 109-139 ; math.NT/0112077.}
  \ref{McInnes}{McInnes,B. {\it APS instability and the topology of the brane
  world}, hep-th/0401035.}
  \ref{BHS}{Brevik,I, Herikstad,R. and Skriudalen,S. {\it Entropy Bound for the
  TM Electromagnetic Field in the Half Einstein Universe}; hep-th/0508123.}
  \ref{BandO}{Brevik,I. and Owe,C.  \prD{55}{4689}{1997}.}
  \ref{Kenn}{Kennedy,G. Thesis. University of Manchester 1978.}
  \ref{KandU}{Kennedy,G. and Unwin S. \jpa{12}{L253}{1980}.}
  \ref{BandO1}{Bayin,S.S.and Ozcan,M.
  \prD{48}{2806}{1993}; \prD{49}{5313}{1994}.}
  \ref{Chang}{Chang, P., {\it Quantum Field Theory on Regular Polytopes}.
   Thesis. University of Manchester, 1993.}
  \ref{Barnesa}{Barnes,E.W. {\it Trans. Camb. Phil. Soc.} {\bf 19} (1903) 374.}
  \ref{Barnesb}{Barnes,E.W. {\it Trans. Camb. Phil. Soc.}
  {\bf 19} (1903) 426.}
  \ref{Stanley1}{Stanley,R.P. \joa {49Hilf}{1977}{134}.}
  \ref{Stanley}{Stanley,R.P. \bams {1}{1979}{475}.}
  \ref{Hurley}{Hurley,A.C. \pcps {47}{1951}{51}.}
  \ref{IandK}{Iwasaki,I. and Katase,K. {\it Proc.Japan Acad. Ser} {\bf A55}
  (1979) 141.}
  \ref{IandT}{Ikeda,A. and Taniguchi,Y. {\it Osaka J. Math.} {\bf 15} (1978)
  515.}
  \ref{GandM}{Gallot,S. and Meyer,D. \jmpa{54}{1975}{259}.}
  \ref{Flatto}{Flatto,L. {\it Enseign. Math.} {\bf 24} (1978) 237.}
  \ref{OandT}{Orlik,P and Terao,H. {\it Arrangements of Hyperplanes},
  Grundlehren der Math. Wiss. {\bf 300}, (Springer--Verlag, 1992).}
  \ref{Shepler}{Shepler,A.V. \joa{220}{1999}{314}.}
  \ref{SandT}{Solomon,L. and Terao,H. \cmh {73}{1998}{237}.}
  \ref{Vass}{Vassilevich, D.V. \plb {348}{1995}39.}
  \ref{Vass2}{Vassilevich, D.V. \jmp {36}{1995}3174.}
  \ref{CandH}{Camporesi,R. and Higuchi,A. {\it J.Geom. and Physics}
  {\bf 15} (1994) 57.}
  \ref{Solomon2}{Solomon,L. \tams{113}{1964}{274}.}
  \ref{Solomon}{Solomon,L. {\it Nagoya Math. J.} {\bf 22} (1963) 57.}
  \ref{Obukhov}{Obukhov,Yu.N. \pl{109B}{1982}{195}.}
  \ref{BGH}{Bernasconi,F., Graf,G.M. and Hasler,D. {\it The heat kernel
  expansion for the electromagnetic field in a cavity}; math-ph/0302035.}
  \ref{Baltes}{Baltes,H.P. \prA {6}{1972}{2252}.}
  \ref{BaandH}{Baltes.H.P and Hilf,E.R. {\it Spectra of Finite Systems}
  (Bibliographisches Institut, Mannheim, 1976).}
  \ref{Ray}{Ray,D.B. \aim{4}{1970}{109}.}
  \ref{Hirzebruch}{Hirzebruch,F. {\it Topological methods in algebraic
  geometry} (Springer-- Verlag,\break  Berlin, 1978). }
  \ref{BBG}{Bla\v{z}i\'c,N., Bokan,N. and Gilkey, P.B. {\it Ind.J.Pure and
  Appl.Math.} {\bf 23} (1992) 103.}
  \ref{WandWi}{Weck,N. and Witsch,K.J. {\it Math.Meth.Appl.Sci.} {\bf 17}
  (1994) 1017.}
  \ref{Norlund}{N\"orlund,N.E. \am{43}{1922}{121}.}
  \ref{Duff}{Duff,G.F.D. \aom{56}{1952}{115}.}
  \ref{DandS}{Duff,G.F.D. and Spencer,D.C. \aom{45}{1951}{128}.}
  \ref{BGM}{Berger, M., Gauduchon, P. and Mazet, E. {\it Lect.Notes.Math.}
  {\bf 194} (1971) 1. }
  \ref{Patodi}{Patodi,V.K. \jdg{5}{1971}{233}.}
  \ref{GandS}{G\"unther,P. and Schimming,R. \jdg{12}{1977}{599}.}
  \ref{MandS}{McKean,H.P. and Singer,I.M. \jdg{1}{1967}{43}.}
  \ref{Conner}{Conner,P.E. {\it Mem.Am.Math.Soc.} {\bf 20} (1956).}
  \ref{Gilkey2}{Gilkey,P.B. \aim {15}{1975}{334}.}
  \ref{MandP}{Moss,I.G. and Poletti,S.J. \plb{333}{1994}{326}.}
  \ref{BKD}{Bordag,M., Kirsten,K. and Dowker,J.S. \cmp{182}{1996}{371}.}
  \ref{RandO}{Rubin,M.A. and Ordonez,C. \jmp{25}{1984}{2888}.}
  \ref{BaandD}{Balian,R. and Duplantier,B. \aop {112}{1978}{165}.}
  \ref{Kennedy2}{Kennedy,G. \aop{138}{1982}{353}.}
  \ref{DandKi2}{Dowker,J.S. and Kirsten, K. {\it Analysis and Appl.}
 {\bf 3} (2005) 45.}
  \ref{Dow40}{Dowker,J.S. \cqg{23}{2006}{1}.}
  \ref{BandHe}{Br\"uning,J. and Heintze,E. {\it Duke Math.J.} {\bf 51} (1984)
   959.}
  \ref{Dowl}{Dowker,J.S. {\it Functional determinants on M\"obius corners};
    Proceedings, `Quantum field theory under
    the influence of external conditions', 111-121,Leipzig 1995.}
  \ref{Dowqg}{Dowker,J.S. in {\it Quantum Gravity}, edited by
  S. C. Christensen (Hilger, Bristol, 1984).}
  \ref{Dowit}{Dowker,J.S. \jpa{11}{1978}{347}.}
  \ref{Kane}{Kane,R. {\it Reflection Groups and Invariant Theory} (Springer,
  New York, 2001).}
  \ref{Sturmfels}{Sturmfels,B. {\it Algorithms in Invariant Theory}
  (Springer, Vienna, 1993).}
  \ref{Bourbaki}{Bourbaki,N. {\it Groupes et Alg\`ebres de Lie}  Chap.III, IV
  (Hermann, Paris, 1968).}
  \ref{SandTy}{Schwarz,A.S. and Tyupkin, Yu.S. \np{242}{1984}{436}.}
  \ref{Reuter}{Reuter,M. \prD{37}{1988}{1456}.}
  \ref{EGH}{Eguchi,T. Gilkey,P.B. and Hanson,A.J. \prp{66}{1980}{213}.}
  \ref{DandCh}{Dowker,J.S. and Chang,Peter, \prD{46}{1992}{3458}.}
  \ref{APS}{Atiyah M., Patodi and Singer,I.\mpcps{77}{1975}{43}.}
  \ref{Donnelly}{Donnelly.H. {\it Indiana U. Math.J.} {\bf 27} (1978) 889.}
  \ref{Katase}{Katase,K. {\it Proc.Jap.Acad.} {\bf 57} (1981) 233.}
  \ref{Gilkey3}{Gilkey,P.B.\invm{76}{1984}{309}.}
  \ref{Degeratu}{Degeratu.A. {\it Eta--Invariants and Molien Series for
  Unimodular Groups}, Thesis MIT, 2001.}
  \ref{Seeley}{Seeley,R. \ijmp {A\bf18}{2003}{2197}.}
  \ref{Seeley2}{Seeley,R. .}
  \ref{melrose}{Melrose}
  \ref{berard}{B\'erard,P.}
  \ref{gromes}{Gromes,D.}
  \ref{Ivrii}{Ivrii}
  \ref{DandW}{Douglas,R.G. and Wojciekowski,K.P. \cmp{142}{1991}{139}.}
  \ref{Dai}{Dai,X. \tams{354}{2001}{107}.}
  \ref{Kuznecov}{Kuznecov}
  \ref{DandG}{Duistermaat and Guillemin.}
  \ref{PTL}{Pham The Lai}
\end{putreferences}

\bye